\documentclass[usenatbib]{mnras}
\usepackage{graphicx}
\usepackage{pdflscape}
\usepackage{newtxtext,newtxmath}
\usepackage{multirow} 
\usepackage{booktabs} 
\usepackage[T1]{fontenc}
\usepackage{ae,aecompl}
\usepackage[dvipsnames]{xcolor}

\title[Magnetic field and chromospheric activity evolution of HD\,75332]{Magnetic field and chromospheric activity evolution of HD\,75332: \\ a rapid magnetic cycle in an F star without a hot Jupiter}

\author[E. L. Brown et al.]{E. L. Brown$^{1}$\thanks{E-mail: emma.brown@usq.edu.au},
S. C. Marsden$^{1}$,
M. W. Mengel$^{1}$,
S. V. Jeffers$^{2}$,
I. Millburn$^{1}$,
M. Mittag$^{3}$,
\newauthor  
P. Petit$^4$,
A. A. Vidotto$^5$,
J. Morin$^6$, 
V. See$^7$,
M. Jardine$^8$,
J.N. Gonz\'alez-P\'erez$^3$ and
\newauthor
the BCool Collaboration
\\
$^{1}$University of Southern Queensland, Centre for Astrophysics, Toowoomba, QLD, 4350, Australia\\
$^{2}$Max Planck Institut for Solar System Research, Justus-von-Liebig-Weg 3, 37077 G\"ottingen, Germany\\
$^{3}$Hamburger Sternwarte, Universit\"at Hamburg, Gojenbergsweg 112, 21029 Hamburg, Germany\\
$^4$IRAP (Institut de Recherche en Astrophysique et Plan\'etologie), Universit\'e de Toulouse, CNRS, CNES, UPS, 14 Avenue Edouard Belin, 31400, Toulouse, France\\
$^5$School of Physics, Trinity College Dublin, The University of Dublin, College Green, Dublin-2, Ireland\\
$^6$LUPM, Universit\'e de Montpellier, CNRS, Place Eug\'ene Bataillon, F-34095 Montpellier, France\\
$^7$University of Exeter, Department of Physics \& Astronomy, Stoker Road, Devon, Exeter, EX4 4QL, UK\\
$^8$SUPA, School of Physics and Astronomy, University of St Andrews, North Haugh, St Andrews, KY16 9SS, UK}

\date{Accepted 2020 December 7. Received 2020 November 29; in original form 2020 June 11.}

\pubyear{2020}

\begin{document}
\label{firstpage}
\pagerange{\pageref{firstpage}--\pageref{lastpage}}
\maketitle

\begin{abstract}
Studying cool star magnetic activity gives an important insight into the stellar dynamo and its relationship with stellar properties, as well as allowing us to place the Sun’s magnetism in the context of other stars. Only 61 Cyg A (K5V) and $\tau$\,Boo (F8V) are currently known to have magnetic cycles like the Sun's, where the large-scale magnetic field polarity reverses in phase with the star's chromospheric activity cycles. $\tau$\,Boo has a rapid $\sim$240\,d magnetic cycle, and it is not yet clear whether this is related to the star’s thin convection zone or if the dynamo is accelerated by interactions between $\tau$\,Boo and its hot Jupiter. To shed light on this, we studied the magnetic activity of HD\,75332 (F7V) which has similar physical properties to $\tau$\,Boo and does not appear to host a hot Jupiter. We characterized its long term chromospheric activity variability over 53\,yrs and used Zeeman Doppler Imaging to reconstruct the large-scale surface magnetic field for 12 epochs between 2007 and 2019.  Although we observe only one reversal of the large-scale magnetic dipole, our results suggest that HD\,75332 has a rapid $\sim$1.06\,yr solar-like magnetic cycle where the magnetic field evolves in phase with its chromospheric activity. If a solar-like cycle is present, reversals of the large-scale radial field polarity are expected to occur at around activity cycle maxima. This would be similar to the rapid magnetic cycle observed for $\tau$\,Boo, suggesting that rapid magnetic cycles may be intrinsic to late-F stars and related to their shallow convection zones.
\end{abstract}


\section{Background}
The Sun's $\sim$22 year magnetic cycle is characterized by the reversal of the global magnetic field every $\sim$11 years. Between polarity reversals the magnetic field is wound up from a dipolar, poloidal structure to a complex, higher-order multipolar structure with an increasing toroidal component, reflecting an underlying cyclic dynamo generated by internal rotational shear. The poloidal field is then regenerated from the toroidal field through cyclonic convection or the Babcock-Leighton mechanism \citep{Charbonneau2005}. The magnetic dynamo manifests at the solar surface in the formation and evolution of bright chromospheric plages and dark photospheric sunspots throughout an $\sim$11 year `activity cycle'. The activity cycle modulates chromospheric emissions (e.g. in the cores of the \ion{Ca}{ii} H\&K spectral lines), high energy emissions originating from the transition region and corona at EUV and X-ray wavelengths, and the solar wind that impact on our space environment. 

Activity cycles analogous to the solar activity cycle have been detected for many cool stars  \citep{Wilson1978,Baliunas1995,Lehtinen2016,BoroSaikia2018}. Magnetic heating in stellar chromospheres causes emission in the cores of the \ion{Ca}{ii} H\&K spectral lines (among others), similar to what is observed in the Sun \citep{Eberhard1913}. The Mount Wilson (MW) HK Project monitored chromospheric \ion{Ca}{ii} H\&K emissions for over 2000 cool stars between 1966 and 2001. The MW S-index is the ratio of the fluxes in the cores of the \ion{Ca}{ii} H\&K spectral lines to the continuum flux \citep{Wilson1978, Baliunas1995}. \citet{Baliunas1995} were the first to detect solar-like chromospheric S-index cycles for 45 MW stars, having spectral types ranging from F to K and showing cycle periods between 2 and 30 years. They also identified stars with significant S-index variability but no apparent periodicity, as well as stars that had `flat' S-index profiles. Follow-up studies continue to expand the database of stars with detected chromospheric activity cycles \citep*{Hall2007,BoroSaikia2018} and examine how stellar properties relate to activity cycle periods \citep*{Bohm2007,Brandenburg2017,BoroSaikia2018} and shapes \citep{Willamo2020}. For a number of targets that were initially classified  as having no apparent S-index periodicity \citep{Baliunas1995}, such as the F-type stars HD\,75332, HD\,16673 and HD\,100563, reanalysis has indicated very low amplitude and rapid ($<2$\,yr) chromospheric activity cycles \citep{Mittag2019}. Such rapid cycles appear to be common among late-F stars and may be driven by different mechanisms compared to the longer and more pronounced chromospheric activity cycles detected in cooler, less massive stars like the Sun.  

It is not yet clear whether solar-like magnetic cycles are as common among cool stars as solar-like chromospheric activity cycles. A key feature of the solar magnetic cycle is the reversal of the large-scale magnetic field polarity in-phase with chromospheric activity cycles. Observing such structural changes in stellar magnetic fields is made possible by the technique of Zeeman Doppler Imaging \citep[ZDI,][]{Semel1989}, a tomographic imaging technique that reconstructs the large-scale surface magnetic fields of stars using high-resolution spectropolarimetric data. Around 20 stars have been observed using ZDI over multiple epochs, many by the BCool\footnote{https://bcool.irap.omp.eu} collaboration. For most stars, magnetic field variability has been observed, but without any clear periodicity \citep{Morgenthaler2011,Jeffers2014,BoroSaikia2015}. For the F9V star HD\,78366, \citet{Morgenthaler2011} observed two reversals of the large-scale field polarity, suggesting a possible $\leq$3\,yr magnetic cycle. However the magnetic cycle does not correspond to the star's detected chromospheric activity cycles \citep[5.9 yr and 12.2 yr, ][]{Baliunas1995}. Only 61 Cyg A (K5V) \citep{BoroSaikia2016,BoroSaikia2018a} and $\tau$\,Boo ($\tau$\,Boo, F8V) \citep{Donati2008,Fares2009,Fares2013,Mengel2016,Jeffers2018} are currently known to have solar-like magnetic cycles where the large-scale field polarity reverses in phase with chromospheric activity cycles. 

$\tau$\,Boo has a rapid, $\sim$240\,d magnetic cycle \citep{Mengel2016,Jeffers2018}, while 61 Cyg A has a $\sim$14\,yr magnetic cycle \citep{BoroSaikia2016} more similar in length to the Sun's $\sim$22\,yr magnetic cycle. $\tau$\,Boo also hosts a hot Jupiter (HJ) with a mass of $\sim 6\,M_{Jupiter}$ orbiting at 0.049\,AU \citep{Borsa2015}, which has been speculated to play a role in generating the rapid stellar magnetic cycle \citep{Fares2009,Vidotto2012}. Since short chromospheric activity cycles like that observed for $\tau$\,Boo appear to be common among F-type stars \citep{Baliunas1997,Metcalfe2010,Mittag2019}, it is possible that rapid magnetic cycles are also common. Indeed, asteroseismic observations have indicated short-period modulations related to internal magnetic activity in F-type stars such as HD\,49933 \citep[F5V, $M=1.17\pm0.1M_\odot$, ][]{Michel2008,Garcia2010}. If rapid magnetic cycles are common among F-type stars, they would be attractive targets to study stellar magnetic cycles within a short time-frame. 

In this study we combined $\sim53$ years of chromospheric activity data with multi-epoch spectropolarimetric observations to study the chromospheric variability and magnetic field evolution of the F7V star HD\,75332. HD\,75332 is a solar-type star with similar physical properties to $\tau$\,Boo, though it does not appear to host a HJ. This makes it an ideal comparison star to probe the possible nature of any star-planet interaction that may contribute to the rapid magnetic cycle observed for $\tau$\,Boo. The key stellar parameters for HD\,75332 and $\tau$\,Boo are shown in Table \ref{tab:StellarParam}. Based on $\sim25$ years of MW observations spanning the years 1966 to 1991, \citet{Baliunas1995} reported significant variability in the chromospheric S-index for HD\,75332, but found no clear activity cycles with potential periods between 2 and 30 years. Using an additional $\sim\,10$ years of MW data, \citet{Olspert2018} detected a $22.8\,\pm\,0.47$\,yr S-index cycle. Also, \citet{Mittag2019} detected a rapid $179.9\ \pm\ 1.0$\,d chromospheric activity cycle using high-cadence spectroscopic observations taken between 2013 and 2018. HD\,75332 was included in the BCool project's magnetic snapshot survey \citep{marsden2014} but the large-scale magnetic field geometry has not been previously reconstructed using ZDI.

\begin{table}\label{tab:StellarParam}
\caption{Key parameters for HD\,75332 and $\tau$\,Boo.} 
\begin{tabular}{ccccc}
\toprule
Parameter &     & HD\,75332 & $\tau$\,Boo & Reference \\ \midrule
T$_{\textrm{eff}}$  & (K) & 6258 $\pm$ 44  & 6399 $\pm$ 45 & 1, 2 \\ [+0.7mm]
log(M/H)  & - &  $+0.05\pm0.03$ &  $+0.25\pm\ 0.02$ & 1 \\ [+0.7mm]
log g     & (cm s$^{-1}$) & 4.34 $\pm$ 0.03   & 4.27 $\pm\ 0.06$ & 3, 2  \\ [+0.7mm]
Age 	& (Gyr)	&   1.88 $^{+0.72}_{-0.92}$   &   0.9 $\pm$ 0.5 &      3, 2   \\ [+0.7mm]
Mass	& ($M_{\sun}$)	&   1.211 $^{+0.026}_{-0.020}$   &   1.39 $\pm$ 0.25 &    3, 2      \\ [+0.7mm]
Radius	& ($R_{\sun}$)	&   1.24 $^{+0.05}_{-0.03}$   &   1.42 $\pm$ 0.08  &    3, 2   \\ [+0.7mm]
B-V 	&  -  &	0.549 $\pm\ 0.005$ & 0.508 $\pm\ 0.001$	& 4  \\ [+0.7mm]
$B_{l}$	&	(G)			& 	$-8.1\pm2.6$		& 3.2 $\pm\ 0.5$ & 5 \\[+0.7mm]
$\langle{S}\rangle$		&	-		&	$0.279\pm0.010$	& 0.202 &	6, 7		\\[+0.7mm]
$v\sin{i}$	& (km s$^{-1}$)     & 9 $\pm$\ 0.5   & 14.27 $\pm$\ 0.06 & 1, 2         \\ [+0.7mm]
RV	& (km s$^{-1}$)     & +4.4 $\pm\ 0.1$  & -16.94 $\pm\ 0.35$ &  8, 9        \\ [+0.7mm]
P$_{\textrm{rot}}$		& (d)		& 3.56 $^{+0.11}_{-0.14}$		& 3.1 $\pm$ 0.1	&  8, 10 \\ [+0.7mm]
$R_o$ & - & 0.80 & 1.82 & 8, 5  \\ [+1mm]
\bottomrule
\end{tabular}\\
$^1$\,\citet{Valenti2005}, $^2$\,\citet{Borsa2015}, $^3$\,\citet{Takeda_2007}, $^4$\,\citet{Perryman1997}, $^5$\,\citet{marsden2014}, $^6$\,\citet{Mittag2019}, $^7$\,\citet{Wright2004}, $^8$\,this study, $^9$\,\citet{Gaia2018}, $^{10}$\,\citet{Mengel2016}
\end{table}
\section{Observations}

\subsection{NARVAL spectropolarimetric observations}

High-resolution spectropolarimetric observations of HD\,75332 were obtained using the NARVAL\footnote{http://www.ast.obs-mip.fr/projets/narval/v1/} spectropolarimeter \citep{Auri2003} coupled with the 2\,m T\'elescope Bernard Lyot at Observatoire Pic du Midi, France. NARVAL is a twin of the ESPaDOnS spectropolarimeter at the Canada France Hawaii Telescope, having a spectral coverage of 3700  to 10480 Angstrom with a resolution of $\sim65000$. A total of 95 Stokes {\it{V}} (circularly polarized) observations were collected between November 2006 and January 2019.  Details of individual observations are shown in Table \ref{tab:NARVALobsdetails}, and all NARVAL data presented here are publicly available through the PolarBase\footnote{http://polarbase.irap.omp.eu/} data base \citep{Petit2014}.   Each Stokes {\it{V}} observation was calculated from a series of four Stokes {\it{I}} (intensity) sub-exposures. Each sub-exposure measured intensity in two orthogonal polarization states, with the positions of the orthogonally polarized beams switched between sub-exposures by rotating the retarding rhombs of the polarimeter. 

\subsection{Reduction and calibration}

Observations were reduced and calibrated using the software {\sc{libre-esprit}}, which is based on {\sc{esprit}} (\'Echelle Spectra Reduction: an Interactive Tool) described in \citet{Donati1997}. The program automatically calibrates the Stokes {\it{I}} sub-exposures using a series of calibration exposures taken during the observing runs. Each series of sub-exposures is added to produce a `mean' Stokes {\it{I}} spectrum \citep{Donati1997}. The Stokes {\it{V}} spectrum is determined by dividing sub-exposures with orthogonal polarization states, a technique that removes spurious polarization signals. A null spectrum is also obtained by dividing spectra with identical polarization states, providing a measure of noise and a diagnosis of the reliability of the polarimetric measurement.

\subsection{Least Squares Deconvolution}

{\sc{libre-esprit}} was used to automatically  extract the polarimetric signal from the calibrated spectra with the technique Least-Squares Deconvolution \citep[LSD,][]{Donati1997}, which creates an average spectral line profile from the thousands of lines within a spectrum. This ensures a high signal to noise ratio (SNR) in the LSD spectral line profiles, which is particularly important for low activity, solar-type stars such as HD 75332.  

LSD requires the generation of a spectral line mask using a model stellar atmosphere for a star with similar properties to the target. We used a line mask generated by \cite{marsden2014} using the Vienna Atomic Line Database \citep[VALD, ][]{Kupka2000}, with $T_{eff}=6250\,\textrm{K}$, $\log{g}=4.5\,\textrm{cm s}^{-2}$ and $\log\textrm{(M/H)}=0.00$. The line mask excludes very strong features such as the \ion{Ca}{ii} H\&K spectral lines and is deconvolved from the Stokes {\it{V}} spectrum to determine the polarization signature. SNRs for the Stokes {\it{V}} LSD line profiles are included in Table \ref{tab:NARVALobsdetails}. 
\section{Chromospheric activity and variability}

\subsection{Chromospheric S-index}

The S-index is the most commonly used index for stellar chromospheric activity, and decades-worth of historical data are available for the analysis of long term activity variability. We combined  archival S-index observations from the Mount Wilson Observatory (MWO) with derived S-indices from our NARVAL observations and S-index observations obtained using the TIGRE telescope. 

\subsubsection{MWO S-indices}

The MW HK Project, carried out at MWO, was a collaborative, multi-decade survey of the chromospheric activity of cool stars \citep*{Wilson1978,Vaughan1978,Duncan1991,Baliunas1995}. It culminated in a database of MW S-index observations for almost 2300 stars, which is openly available to the research community. The MW S-index is based on the ratio of the fluxes in the cores of the chromospheric H and K spectral lines to the fluxes in the continuum either side of the H and K lines.  S-index observations carried out with different instruments can be converted to the MWO scale using Equation \ref{eq:Sindex} from \citet{marsden2014}, which includes the calibration coefficients a, b, c, d and e, that depend on the observing instrument. 

\begin{equation}\label{eq:Sindex}
    S_{MW}=\frac{a F_H + b F_K}{c F_{R_{HK}} + d F_{V_{HK}}} +e
\end{equation}
where $F_{H}$ and $F_{K}$ are the fluxes in two triangular bandpasses centred on the cores of the \ion{Ca}{ii} H and K lines (3968.469\,{\AA} and 3933.663\,{\AA})  with widths of 2.18\,{\AA} (at the base), and $F_{R_{HK}}$ and $F_{V_{HK}}$ are the fluxes in two rectangular 20\,{\AA} bandpasses centered on the continuum either side of the H\&K lines at 3901.07\,{\AA} and 4001.07\,{\AA}. 

HD\,75332 was monitored as part of the MW Project between 1966 and 2011, providing 2380 S-index observations which we have included in our study. 

\subsubsection{NARVAL S-indices}

We calculated S-indices from our reduced and calibrated NARVAL Stokes {\it{I}} observations using a similar process to \citet{Wright2004} and \citet{marsden2014}. We used an auto-extraction pipeline in Python to identify the orders containing the \ion{Ca}{ii} H\&K spectral lines and the required continuum bandpasses, as well as remove any overlap between the orders.  Wavelengths were corrected for the stellar radial velocity (RV), which we estimated by fitting a pseudo-Voigt profile  (convolution of a Gaussian and Lorentzian) to each Stokes {\it{I}} LSD line profile and taking the central velocity as the RV. We then extracted S-indices using Equation \ref{eq:Sindex} where the coefficients a, b, c, d and e are derived in \cite{marsden2014} to calibrate NARVAL observations to the MW S-index scale. We derived a mean S-index of $0.272\,\pm\,0.008$ for the NARVAL data. Our derived mean S-indices for each series of four Stokes {\it{I}} sub-exposures are included in Table \ref{tab:NARVALobsdetails}. 

\subsubsection{TIGRE S-indices}

We made use of 155 S-index observations of HD\,75332 that were obtained between 2013 and 2019 using the TIGRE telescope, which is located at La Luz Observatory near Guanajuato, Mexico, and is described in \citet{schmitt2014}. This data included 128 observations previously published by \citet{Mittag2019}. All TIGRE observations were previously calibrated to the MW S-index scale as described by \citet{Mittag2016}. Further information on the derivation of S-indices from TIGRE observations can be found in \citet{Mittag2016} and \citet{Mittag2019}.  

\subsubsection{Combined S-index time series}

\begin{figure*}
    \centering
    \includegraphics{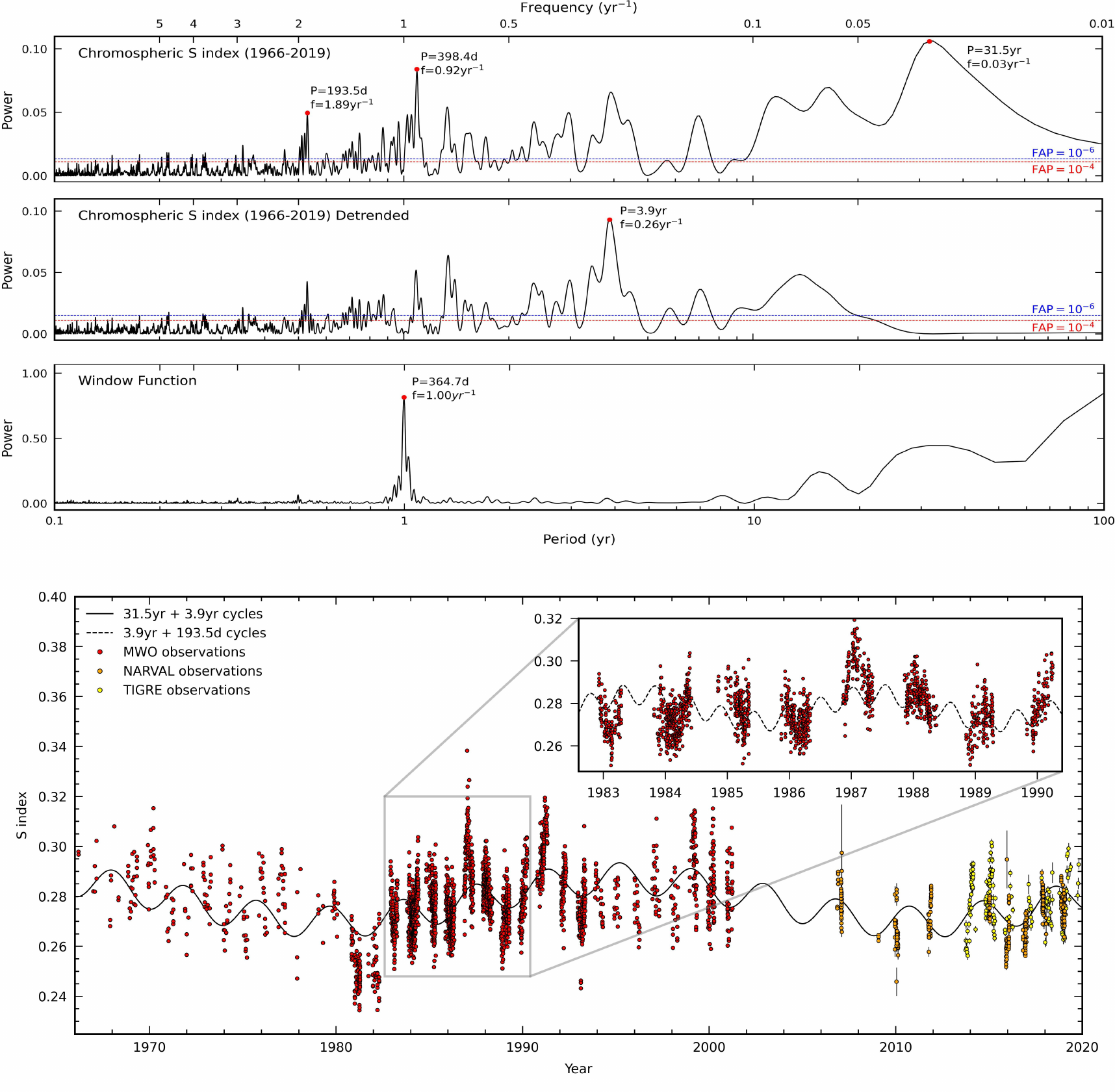}
    \caption{Top: Power spectrum for the chromospheric S-index, computed using the GLS periodogram and including all available S-index data from NARVAL, MWO and TIGRE.
    Second: Power spectrum for the de-trended S-index data. The 31.5\,yr cycle has been removed.
    Third: Power spectrum for the Window Function.
    Bottom: Time series of S-index data showing the fit of the detected cycles to observations. The main chart shows the fit of 31.5\,yr and 3.9\,yr cycles to the data, and the inset axis shows the fit of the 3.9\,yr and 193.5\,d cycles to the data. The x and y scales for the inset axis have been enlarged for clarity. The error bars shown for the NARVAL data were determined by propagating the uncertainties computed during the reduction process for each spectral bin of the normalized Stokes {\it{I}} spectra through Equation \ref{eq:Sindex}. Error bars for the TIGRE data were computed previously by \citet{Mittag2016} and \citet{Mittag2019}. No error measures are available for the MWO data.}
    \label{fig:Sindex}
\end{figure*}

Figure \ref{fig:Sindex} shows the time series of S-index data from MWO, NARVAL and TIGRE. For the combined data we derived a mean S-index of $0.278\pm0.016$, which is consistent with the values obtained by \citet{Baliunas1997} ($\langle{S}\rangle=0.279$) and \citet{Mittag2019} ($\langle{S}\rangle=0.279\pm 0.010$).

\subsection{Period search}\label{subsec:S-index period search}

\subsubsection{GLS periodogram}
To search for periodic variations in the S-index we applied the Generalized Lomb-Scargle (GLS) periodogram derived by \citet{Zechmeister2009}. The GLS periodogram is similar to the standard Lomb-Scargle method \citep{Lomb1976,Scargle1982}, which is equivalent to least-squares fitting of sine waves, though it also allows for a floating (non-zero) mean. We used the Python code by \citet{Zechmeister2009} to compute a power spectrum for the S-index, where cycles present in the data should correspond to a larger periodogram power. 

We assessed the significance of power spectrum peaks with False Alarm Probabilities (FAPs) estimated using two methods. Both methods incorporated bootstrap re-sampling to generate a large number (N) of re-sampled data sets; this involved retaining the timestamps of the S-index observations and assigning random S-index values (with replacement) to each time stamp. We then used the GLS periodogram to find the highest-power peak for each re-sampled data set. For our first estimate of the FAP, we used the maximum peak heights from all N data sets as an estimate of the peak-height probability distribution. The FAP for any signal detected in the real S-index data was calculated as the percentage of re-sampled data sets for which a peak with the same or a greater power was detected \citep{Kuerster1997,Robertson2013,Reinhold2017,Benatti2020}. This method (method 1) allowed FAPs to be calculated down to $\frac{1}{N}$. As a second, more statistically sound estimate of the FAP (method 2), we divided our re-sampled data sets into 100 subsets, providing 100 realizations of the peak-height probability distribution. For any signal in the real S-index data, we calculated the final FAP by taking the average of the FAPs from the 100 probability distributions. This allowed FAPs to be estimated accurately down to $\frac{1}{100N}$. 

Applying the GLS periodogram to many data sets is computationally intensive, in particular when searching for high-frequency signals within an extended time-series, or when searching over a wide frequency range. Thus, when searching for a high-frequency rotational signal (Section \ref{subsec:rotmod}) we computed the GLS periodogram for $10^{4}$ re-sampled data sets, which allowed us to estimate the FAP of detected signals down to $10^{-4}$ using method 1 ($\textrm{FAP}_{1}$), or $10^{-2}$ using method 2 ($\textrm{FAP}_{2}$). After removing the rotation period (as described in Section \ref{subsec:rotmod}) we then limited our periodogram searches to frequencies below $10\,\textrm{yr}^{-1}$, and used $10^{6}$ re-sampled data sets to estimate  $\textrm{FAP}_{1}$ down to $10^{-6}$ and $\textrm{FAP}_{2}$ down to $10^{-4}$.

We also computed a Window Function for the S-index data, which characterizes patterns in the timing of observations and can be derived by simply setting all observations to 1. The observed S-index signal is taken to be the product of the real, continuous signal and the Window Function \citep{VanderPlas2018}. We applied the Lomb-Scargle periodogram (no floating mean) to the Window Function to determine any significant signals that may arise in the S-index power spectrum due to observational cadence. 

\subsubsection{Removing rotational modulation}\label{subsec:rotmod}

\begin{figure}
    \centering
    \includegraphics[width=0.85\columnwidth]{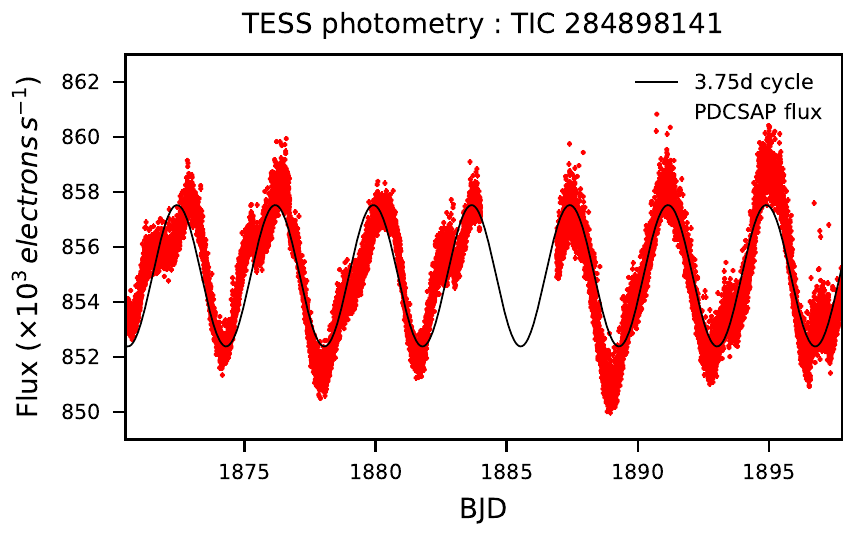}
    \caption{TESS Pre-search Data Conditioned Single Aperture Photometry (PDCSAP) observations of HD\,75332 (red) with a 3.75\,d cycle (black) fitted to the data using the GLS periodogram \citep{Zechmeister2009}. }
    \label{fig:TESSlc}
\end{figure}

The S-index may be modulated by stellar rotation as active regions rotate in and out of view from the observer. This signal should be removed from the S-index data prior to searching for longer periodicities \citep{Olspert2018}. We applied the GLS periodogram to the combined MWO, TIGRE and NARVAL observations and detected two high-frequency signals with $\textrm{FAP}_{1}\,<\,10^{-4}$ and $\textrm{FAP}_{2}<10^{-2}$ (i.e. we did not detect a signal with equal or greater periodogram power in any of our $10^4$ re-sampled data sets). The signals had periods of 3.6849\,$\pm\,0.0001$\,d (hereafter 3.69\,d, amplitude $=0.0028\,\pm\,0.0004$) and 6.6751\,$\pm\,0.0004$\,d (hereafter 6.68\,d, amplitude$\,=\,0.0028\,\pm\,0.0004$). To unambiguously constrain the rotation period we carried out an additional period search using photometric observations from NASA's Transiting Exoplanet Survey Satellite \citep[TESS, ][]{Ricker2015}. TESS observed HD\,75332 (TIC\,284898141) in 2\,min short-cadence integrations from 1 - 28 January 2020. We accessed the light curve from the Mikulski Archive for Space Telescopes (MAST\footnote{https://mast.stsci.edu/portal/Mashup/Clients/Mast/Portal.html}) and it had been pre-processed using the Science Processing Operations Center \citep[SPOC, ][]{Jenkins2016} pipeline to remove long-term trends in the fluxes. We detected a clear photometric period of $3.746\,\pm\,0.002$\,d (Figure \ref{fig:TESSlc}). Therefore we expect that the 3.69\,d signal we detected in the S-index corresponds to the chromospheric rotation period. To remove the rotational signal from the S-index data we fitted a sinusoidal function with a period of 3.69\,d to the data using the GLS periodogram, and then subtracted the function from the S-indices. 

\subsubsection{Activity cycles}

Figure \ref{fig:Sindex} shows the results of the period search after the removal of the rotational signal. For each of the S-index power spectra shown in Figure \ref{fig:Sindex} we have plotted the periodogram powers that correspond to $\textrm{FAP}_{1}=10^{-6}$ and $\textrm{FAP}_{2}=10^{-4}$. All of the detected cycles that we discuss in the following paragraphs have FAPs below these limits. 

The power spectrum shows two well-defined, short-period cycles of $193.5\pm0.2$\,d ($\textrm{amplitude}=0.0048\pm0.0004$) and $398.4\pm0.5$\,d ($\textrm{amplitude}=0.0066\pm0.0004$). Given that the longer of the two cycles is around double the shorter cycle it is possible that these signals are harmonics, in which case the 398.4\,d cycle would be the true signal (having the fundamental frequency). However, we fitted each cycle to the time series data and by visual inspection the 193.5\,d cycle gave a significantly better fit, suggesting that it is a real cycle and not an alias of the $398.4$\,d signal. It should also be noted that a 193.5\,d period is similar to the time-span of observations during each observing season ($\sim180\pm30\,$d), though the strong visual fit of the 193.5\,d model to the S-index data suggests that there is true chromospheric variability on this time-scale. Also, if the signal were related to the observing seasons, we would expect to see the same signal, with a similar periodogram power, in the re-sampled data sets when computing the FAPs. However, none of the re-sampled data sets showed a high-power signal comparable to the power of the 193.5\,d cycle. We also considered the possibility that this short-term variability could be related to the growth and decay of active regions or complexes of activity \citep{Donahue1997}. However, the evolution of active regions and active region complexes would likely show a range of time-scales, amplitudes and phases \citep{Donahue1995}, so we consider it to be unlikely that this could mimic the 193.5\,d signal that prevails throughout many years of observations in the inset of Figure \ref{fig:Sindex}. The $193.5$\,d cycle is within 10\,\% of the $179.9 \pm 1.0$\,d cycle ($\textrm{amplitude}=0.0095$) detected by \citet{Mittag2019}, who also adopted the methods of \citet{Zechmeister2009} for their period search, though they used their own Python code. We applied the \citet{Zechmeister2009} code to the subset of TIGRE data used by \citet{Mittag2019} and detected a cycle with a period of $180.8\pm1.9$\,d ($\textrm{amplitude}=0.0097\pm0.0009$), which is consistent with their result.

The highest-power peak in the power spectrum corresponds to a $31.5 \pm 0.7$\,yr cycle ($\textrm{amplitude}=0.0087\pm0.0004$) but the peak is poorly defined since the S-index data cover less than two cycles of this length. The peak roughly corresponds to a peak in the Window Function power spectrum, so it can not be ruled out that the detected cycle is related to the observational cadence or time-span covered by the data. However, visual inspection of the time series data supports the presence of a low-amplitude, long-period cycle. This would be consistent with the secular decrease in chromospheric \ion{Ca}{ii} H\&K flux reported by \citet{Wilson1978} between 1966 and 1978, and may also be consistent with the long-term S-index cycle detected by \citet{Olspert2018}. Improving our estimate of the length of the cycle relies on further extending our observations to cover multiple cycles.

Several other significant peaks are present in the S-index power spectrum, though they are poorly defined. It is clear also in the time series data that there are additional periodicities present. We de-trended the S-index data by subtracting the detected 31.5\,yr cycle and recomputed the GLS periodogram and the FAPs. The power spectrum for the detrended data shows a significant $3.89 \pm 0.02$\,yr cycle ($\textrm{amplitude}=0.0063\pm0.0004$, $\textrm{FAP}_{1}<10^{-6}$, $\textrm{FAP}_{2}<10^{-4}$, period taken as 3.9\,yr hereafter) which appears to match the time series data well.  

As expected, the power spectrum for the Window Function shows a significant annual peak ($f=1.00 yr^{-1}$) due to the seasonal observations. This does not appear to correspond directly to any significant peaks in the S-index power spectrum. It is also possible that the annual observations could cause a series of aliases in the S-index power spectrum either side of any real S-index signal at frequency intervals of $1.00 yr^{-1}$ \citep{VanderPlas2018}. Therefore, it is possible that the 398.4\,d cycle is an alias of the 193.5\,d cycle, caused by the observation frequency, as the frequency interval is equal to $0.97 yr^{-1}$. 
\section{Magnetic field variability}\label{Reconstruction of large-scale surface magnetic field}

\subsection{Zeeman Doppler Imaging}

The large-scale magnetic field of HD\,75332 was reconstructed from the NARVAL Stokes {\it{V}} observations using ZDI. ZDI was carried out using the program {\sc{zdipy}} described in \citet{Folsom2018a}, which is a Python-based code that mirrors the extensively used, original C implementation of \citet{Donati2006}. 

{\sc{zdipy}} models the surface of a star as a grid of equal-area tiles, each with a magnetic field made up of radial, azimuthal and meridional components. The magnetic field for each tile is expressed as a series of spherical harmonics \citep{Donati2006}, where the order of the spherical harmonic expansion is limited by a sufficiently large value, $l_{max}$, related to the complexity of the magnetic field structure (e.g. $l=1$ for a dipolar structure, $l=2$ for quadrupolar structure, etc.). Model line profiles for each tile are represented as a function of the magnetic field using a Voigt profile as the basis for Stokes {\it{I}} profiles, and using the weak-field approximation to calculate the Stokes {\it{V}} profiles \citep{Donati1997}. Full disk-integrated model line profiles combine profiles across all visible tiles, which are Doppler shifted according to the line-of-sight projection of the rotational velocity of the tile, and scaled according to the projected area of the tile and to account for limb darkening. 

{\sc{zdipy}} iteratively varies the magnetic field and reconstructs model line profiles for each iteration, comparing them to the observed LSD profiles. The optimum model line profiles are those that fit the observed LSD profiles within a desired reduced-$\chi^2$ value. Meanwhile, the magnetic field is reconstructed from the model Stokes {\it{V}} profiles using a principle of maximum entropy image reconstruction, wherein the image content is minimized \citep{Donati2006}.

We organised our Stokes {\it{V}} observations into 21 subsets as shown in Table \ref{tab:NARVALdata}. The epochs listed for each data subset are close to the mid-point of the subset, but an integer multiple of rotations from our first NARVAL observation on 14 November 2006 (HJD 2454054.67411). Given the possibility of a short magnetic cycle occurring in phase with our detected short chromospheric activity cycle, we limited the time span of observations for each subset to less than 1 month and avoided large gaps of multiple stellar rotations between observations. This ensured that our magnetic reconstructions were unlikely to be significantly affected by the evolution of magnetic features.  We tested larger combinations of data that spanned more than one month and gave better phase coverage, but the subsequent reconstructions showed smeared magnetic features suggesting that indeed evolution of the magnetic field was occurring.  12 of the subsets consisted of a sufficient number of observations to reliably recover the large-scale structure of the magnetic field using ZDI. It is desirable to use $\geq6$ observations to derive a useful map of the magnetic field. However, we have also included maps for data sets with 4 observations as the large-scale polarity of the reconstructed field should be reliable; note that any smaller-scale features in these maps should be considered with caution. The minimum size of the ZDI subsets we use here is similar to series' of observations used to model the magnetic fields of $\tau$ Boo \citep{Jeffers2018} and HD78366 \citep{Morgenthaler2011}.  We further discuss the resolution of the ZDI maps in Section \ref{sec:model_line_profiles}.

\begin{table*}
\caption{Summary of NARVAL observations of HD\,75332. For each epoch, columns list the HJD of the first (HJD start) and last (HJD end) Stokes {\it{V}} observations (the HJD at the midpoint of the series of four Stokes {\it{I}} sub-exposures), the number of observations, the rotational cycle of the first ($\phi$ start) and last ($\phi$ end) observations with respect to the ephemeris $HJD=2454054.67411+3.56\phi$, and the number of definite (D) and marginal (M) detections of the polarization signal. A definite detection in the LSD Stokes {\it{V}} profile has a FAP $\leq{10^{-5}}$ and a marginal detection has $10^{-5}<$ FAP $\leq{10^{-3}}$ \citep{Donati1997}.}
\begin{tabular}{llccccccc}
\toprule
 \multirow{2}{*}{Epoch} & & HJD start & HJD end & \multirow{2}{*}{No. obs.} & \multirow{2}{*}{$\phi$ start} & \multirow{2}{*}{$\phi$ end} & \multirow{2}{*}{D} & \multirow{2}{*}{M} \\ 
  & &  (+2454000) & (+2454000) &  &  &  &  & \\ \midrule
2006.89 & Nov 2006  		& 54.67	    & 69.73		& 2	&	0.00	& 4.23		& 2	& 0 \\
2007.09 & Jan - Feb 2007  	& 127.53	& 140.53	& 8	&	20.46	& 24.12		& 4	& 1 \\
2009.08 & Jan 2009  		& 862.61	& 862.61	& 1	&	226.95	& 226.95	& 1	& 0 \\
2009.95 & Dec 2009  		& 1180.73	& 1181.71	& 2	&	316.31	& 316.58	& 0	& 0 \\
2010.01 & 5 Jan 2010  		& 1202.67	& 1202.67	& 1	&	322.47	& 322.47	& 0	& 1 \\
2010.07 & Jan 2010  		& 1215.60	& 1224.58	& 4	&	326.10	& 328.63	& 1	& 2 \\
2010.13 & Feb 2010	        & 1241.56	& 1243.53	& 3	&	333.40	& 333.95	& 0	& 0 \\
2011.77 & Oct 2011	        & 1839.71	& 1848.68	& 4	&	501.41	& 503.93	& 1	& 0 \\
2011.88 & Nov 2011	        & 1874.70	& 1892.70	& 6	&	511.24	& 516.30	& 4	& 0 \\
2015.02 & Jan 2015	        & 3028.63	& 3040.56	& 8	&	835.38	& 838.73	& 7	& 0 \\
2015.94 & Dec 2015	        & 3358.63	& 3374.62	& 4	&	928.08	& 932.57	& 2	& 0 \\
2016.07 & Jan 2016	        & 3408.72	& 3417.57	& 7	&	942.15	& 944.63	& 4	& 1 \\
2016.82 & Oct 2016	        & 3690.60	& 3690.60	& 1	&	1021.33	& 1021.33	& 0	& 1 \\
2016.95 & Dec 2016	        & 3725.68	& 3741.70	& 11 &	1031.18	& 1035.68	& 5	& 1 \\
2017.02 & Jan 2017	        & 3756.59	& 3761.65	& 4	&	1039.87	& 1041.29	& 1	& 1 \\
2017.84 & 1 Nov 2017  		& 4059.57	& 4059.57	& 1	&	1124.97	& 1124.97	& 0	& 0 \\
2017.90 & Nov - Dec 2017  	& 4073.67	& 4093.62	& 11 &	1128.93	& 1134.54	& 4	& 2 \\
2018.06 & Jan 2018  		& 4142.73	& 4142.73	& 1	&	1148.33	& 1148.33	& 0	& 0 \\
2018.87 & Nov 2018  		& 4437.68	& 4439.63	& 2	&	1231.18	& 1231.73	& 1	& 0 \\ 
2018.94 & Dec 2018  		& 4457.72	& 4470.65	& 4	&	1236.81	& 1240.44	& 1	& 2 \\
2019.03 & Jan 2019  		& 4487.64	& 4510.60	& 10&	1245.22	& 1251.67	& 5	& 2 \\ \bottomrule
\end{tabular}
\label{tab:NARVALdata}
\end{table*}

\subsection{Stellar and model parameters}

\subsubsection{Model spectral line profiles}\label{sec:model_line_profiles}
HD\,75332 has an effective temperature close to solar so we approximated the model line profiles based on a solar line model, with $\lambda=650\,\textrm{nm}$, Gaussian width = 2.41\,\textrm{km\,s}$^{-1}$, Lorentzian width = 0.89\,\textrm{km\,s}$^{-1}$, effective Lande factor = 1.195 \citep{Folsom2016,Folsom2018a}, and limb darkening coefficient  $\eta$ = 0.66 \citep[consistent with ][]{Sing2010}. Figure \ref{fig:DI_fit} shows the fit of the Stokes {\it{I}} line model to a sample observation. We used a consistent $l_{max}=12$ for all reconstructed maps, though for several data subsets a lower $l_{max}$ could have been sufficient. For $l_{max}>12$ we did not observe any significant differences in the magnitude or structure of the magnetic field, as next to zero energy was stored in higher order components. Based on the $9\,\pm\,0.5\,\textrm{km s}^{-1}$ projected rotational velocity of HD\,75332 (Table \ref{tab:StellarParam}) the number of spatially resolved elements at the equator is around 8 \citep{Morin2010}, which is equivalent to a longitudinal resolution of $45\degr$. Therefore $l_{max}=12$ is adequate for our reconstructions.

\subsubsection{$\mathcal{\chi}^2$ fit-stopping criterion}\label{subsection:Model fit-stopping criterion}
Our Stokes {\it{V}} line models converged easily to solutions with reduced-$\chi^2$ fits less than 1.0. However, to avoid inappropriately fitting to noise in Stokes {\it{V}} profiles, over-fitting `limits' between $\chi^2=0.95$ and $\chi^2=1$ are typically used. To determine an appropriate $\chi^2$ fit-stopping limit we reconstructed magnetic images with increasingly low $\chi^2$ targets and studied the reduction in image entropy (i.e. growth of image content) and increase in mean field strength \citep{Alvarado_Gomez_2015,Mengel2016}. We observed a rapid decrease in image entropy and increase in magnetic field strength at around $\chi^2=0.8$, suggesting that the model was inappropriately fitting to noise. Since $\chi^2=0.8$ is much lower than typical over-fitting limits, we set $\chi^2_{aim}=0.95$, which is commonly used for magnetic field reconstructions \citep[e.g.][]{Mengel2016}.

\subsubsection{RV, v$\mathcal{\sin}$i and inclination angle}
Stellar RV, $v\sin{i}$ and inclination angle ($i$) were measured from the Stokes {\it{I}} LSD profiles using {\sc{zdipy}}. We applied the technique of \citet{Folsom2018} where models of the stellar surface are computed for a grid of stellar parameters and the parameters that maximize the solution entropy (i.e. minimize the information content) are selected. We derived RVs between $4.3\pm0.1\textrm{\,km\,s}^{-1}$ and $4.5\pm0.1\textrm{\,km\,s}^{-1}$ across epochs. For ZDI modelling we adopted the most common result RV = $4.4\pm{0.1}\textrm{\,km\,s}^{-1}$ for all epochs, since the apparent RV perturbations across epochs were within our measurement uncertainty. This is consistent with the 4.5\,km\,s$^{-1}$ measured by \citet{marsden2014} and the measured values from various other studies in the SIMBAD data base. The measured $v\sin{i}$ across all epochs was consistent within the error bars with the $9\pm0.5\textrm{\,km\,s}^{-1}$ derived by \citet{Valenti2005}, so we adopted this value for ZDI mapping. This value also agrees with that implied by the rotational broadening of the Stokes {\it{I}} spectral line profiles. We measured a best fitting inclination angle of $30\pm5\degr$, which agrees with that implied by the $v\sin{i}$ determined here along with stellar radius and rotation period from Table \ref{tab:StellarParam}. We also derived consistent stellar parameters when applying the method of \citet{Folsom2018} to the Stokes {\it{V}} profiles.

\subsubsection{Equatorial rotation period and differential rotation}\label{sec:diffrot}

No previous studies have measured differential rotation (DR) for HD\,75332, but it is a common feature of main-sequence F-type stars \citep{reiners2007}. {\sc{zdipy}} incorporates a solar-like DR law (Equation \ref{eq:solarDR}), which assumes that the equator rotates at a higher angular velocity compared to the poles. 
\begin{equation}\label{eq:solarDR}
\Omega(\theta)=\Omega_{eq} - d\Omega \sin^2\theta
\end{equation}
where $\Omega(\theta)$ is the angular rotation rate at a latitude of $\theta$, $\Omega_{eq}$ is the equatorial rotation rate and $d\Omega$ is the rotational shear between the equator and the poles, each measured in rad\,d$^{-1}$. Both the equatorial rotation period and rate of differential rotation can be measured using {\sc{zdipy}} if the same region of the stellar surface is observed over multiple rotations.  For a star that is differentially rotating,  the signatures of magnetic regions will be shifted over consecutive rotations rather than repeating exactly. 
The DR parameters can be derived by applying the method of \citet*{Petit2002}, whereby $P_{rot}$ and $d\Omega$ are varied within a reasonable domain, and the magnetic field is reconstructed with a consistent image entropy for each pair of parameters. When DR is detected, the $\chi^2$ values for all the models converge to a paraboloid, from which the optimum DR parameters can be determined.

\begin{figure*}
\includegraphics[width=0.95\linewidth]{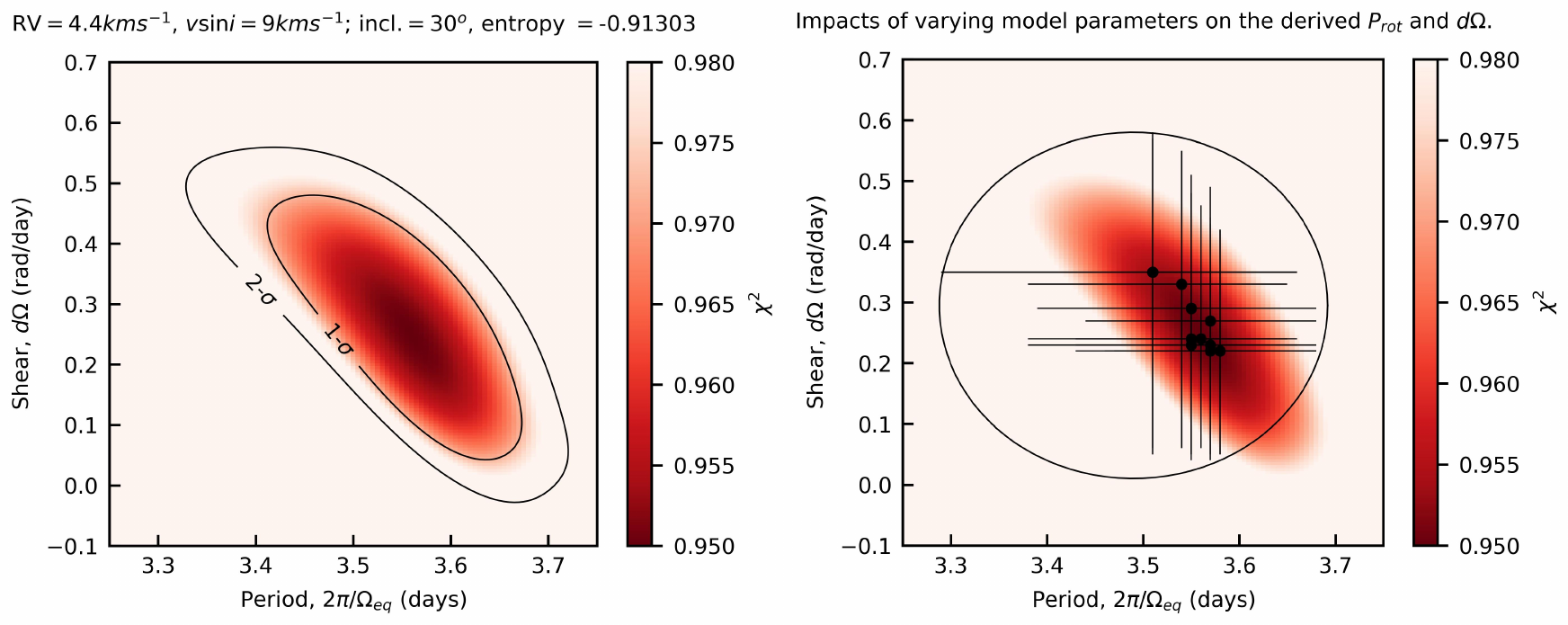}
\caption{Left: Reduced $\chi^2$ as a function of rotational shear (rad d$^{-1}$) and rotation period (days) for HD\,75332, Jan 2015, for the stellar parameters shown. The entropy target was appropriate for reconstruction of the magnetic maps with a minimum $\chi^2$ fit of 0.95. The contours shown are the $1\sigma$ and $2\sigma$ thresholds from the $\chi^2$ minimum. Right: $1\sigma$ uncertainties in the best-fitting differential rotation parameters for Jan 2015 with ZDI model parameters simultaneously varied within the ranges RV\,$\pm\,0.1\textrm{\,km\,s}^{-1}$, $v\sin{i}\pm0.5\textrm{\,km\,s}^{-1}$, inclination angle $\pm\, 5\degr$ and $\chi^2_{aim}\pm0.05$. Each individual point represents the best-fitting DR parameters for a particular combination of model parameters. The fitted ellipse provides a conservative estimate of the possible variations in the measured DR parameters with variations in model parameters.}
\label{fig:DR_2015}
\end{figure*}

We applied this method to all of our ZDI data sets and used the photometric rotation period of 3.75\,d to guide our DR parameter search. The $\chi^2$ landscape for Jan 2015 is shown in Figure \ref{fig:DR_2015}, from which we derived $P_{rot}=3.56\,^{+0.11}_{-0.14}\textrm{\,d}$ and $d\Omega=0.25\,^{+0.23}_{-0.22} \textrm{\,rad\,d}^{-1}$. This suggests an equator-pole lap time of $25^{+184}_{-12}$\,d for HD\,75332, compared to the solar value of $\sim120\,$d. The very large 1$\sigma$ uncertainties in the DR parameters are predominantly caused by the low $v\sin{i}$ of HD\,75332, which limits the spatial resolution of the stellar surface and reduces the prospects for accurately measuring DR \citep{Petit2002}.  We also show in Figure \ref{fig:DR_2015} (right) the impacts of varying the adopted model parameters on the derived DR parameters and their uncertainties. 

Analyses using data from Jan-Feb 2007, Dec 2016 and Jan 2019 all yielded DR parameters that were consistent with the results from Jan 2015 (i.e. within the 1$\sigma$ uncertainty region). For the remaining data sets we were unable to determine reliable estimates for the DR parameters because the $\chi^2$ landscapes did not converge to clear minima, or they suggested solid-body rotation. This was likely due to the low numbers and poor phase coverage of observations, or inconsistent observational cadence, all of which can significantly impact on the measured DR parameters \citep{Petit2002}. We consider the parameters derived for Jan 2015 to be the most reliable since the data is high quality, having the greatest number of definite detections of the magnetic field compared to all other data sets (Table \ref{tab:NARVALdata}), and observations that evenly sample the stellar surface over $\sim4$ rotations. Therefore, for ZDI we have adopted $P_{rot}=3.56$\,d and $d\Omega=0.25\textrm{\,rad\,d}^{-1}$ for all epochs. We also carried out sensitivity testing to determine the possible impacts of the adopted DR parameters (and other model parameters) on the derived magnetic field properties (see Section \ref{section:Sensitivity analysis} for further details).

The adopted DR parameters are consistent with our measured 3.69\,d chromospheric and 3.75\,d photometric rotation periods if we assume that they correspond to the rotation of active regions at latitudes of $\sim30\degr$ and $\sim37\degr$ respectively. The derived rotational shear is also within the range of values measured for similar stars by \citet{Balona2016} ($0\leq{d\Omega}\leq{0.8}\textrm{\,rad\,d}^{-1}$), and is consistent with the expected values from hydrodynamic numerical modelling by \citet{augustson2012} and \citet{Brun2017}. Using equations 24 and 25 from \citet{augustson2012} and a Rossby number determined from the empirical fits of \citet{Noyes1984} (Table \ref{tab:StellarParam}) we expect $0.22\leq{d\Omega_{60}}\leq{0.56}\textrm{\,rad\,d}^{-1}$, where $d\Omega_{60}$ is the rotational shear between the equator and $60\degr$ latitude.  From Figure 14 and Equation 33 of \citet{Brun2017} we derive $d\Omega_{60}\approx0.35 \textrm{\,rad\,d}^{-1}$.

We also carried out a sliding-window S-index period search to track changes in the chromospheric rotation period over time, which can provide an additional lower-limit for the rotational shear \citep{Donahue1995}.  We applied the GLS periodogram to subsets of S-index data within a 100\,d sliding window and adopted a maximum $\textrm{FAP}_{1}$ of $10^{-3}$ to identify significant periodicities (see Section \ref{subsec:S-index period search}). We detected possible variations in the rotational signal between $3.52\,\pm{0.02}\,d$ (amplitude$\,=\,0.0046\pm0.0009$) and $3.84\pm{0.03}\,d$ (amplitude$\,=\,0.0046\pm0.0008$), which may occur over a rapid time-scale ($\sim100$\,d).  Assuming that these rotation periods correspond to active regions located at the equator and the visible pole (i.e. $\Omega_{eq}=1.79\,\pm0.01\,\textrm{rad\,d}^{-1}$ and $\Omega_{\frac{\pi}{2}}=1.64\pm0.02\,\textrm{rad\,d}^{-1}$), they suggest a lower limit for the rotational shear of $0.15\,\pm{0.02}\,rad\,d^{-1}$ (Equation \ref{eq:solarDR}), which equates to a maximum equator-pole lap time of $\sim42^{+6}_{-5}\,d$. This result is within the 1-$\sigma$ range derived for d$\Omega$ (Jan 2015) from ZDI.

\subsection{ZDI results}\label{subsec:ZDIresults}

\begin{figure*}
\includegraphics[width=0.95\linewidth]{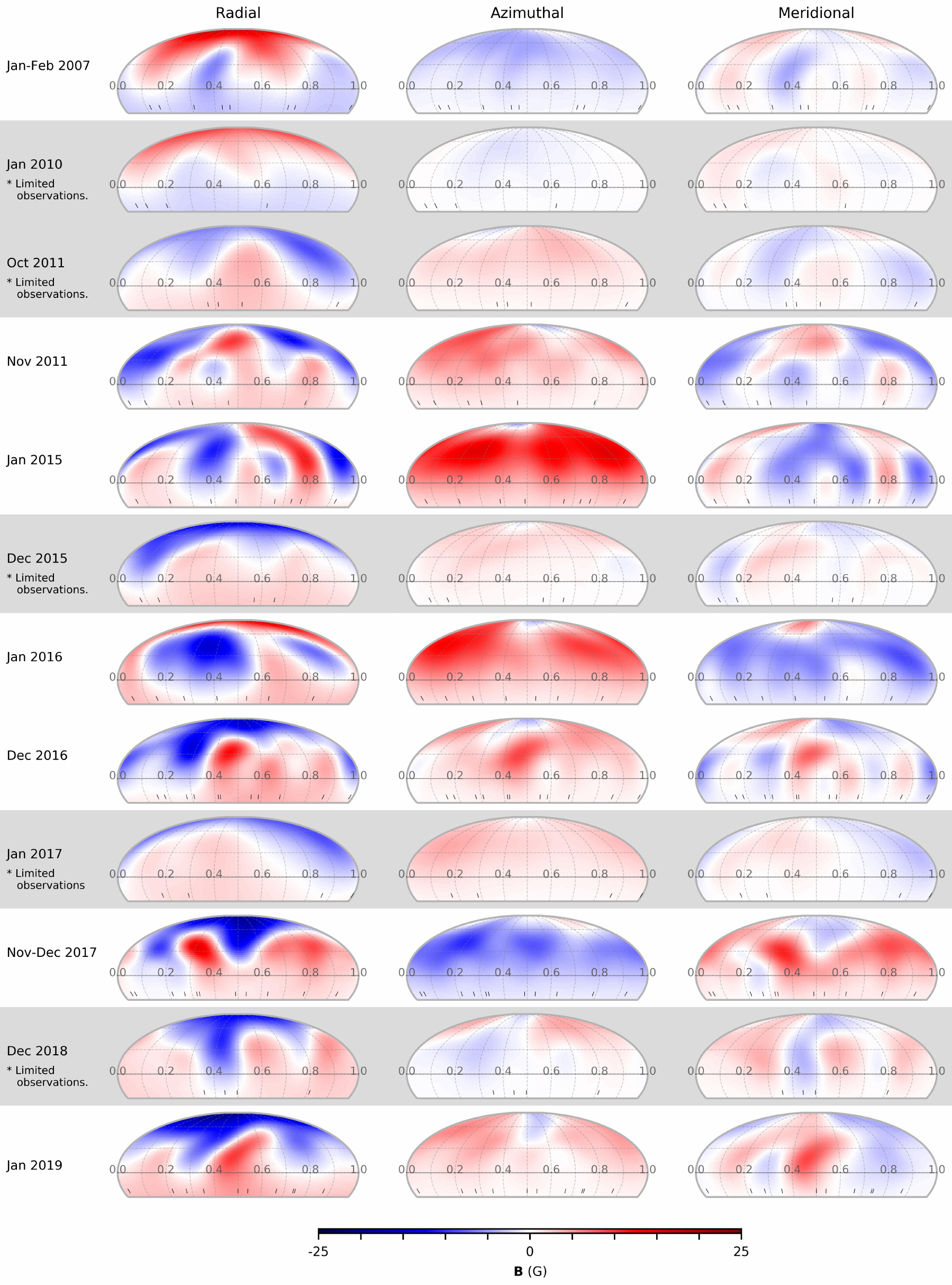}
\caption{Reconstructed magnetic topology of HD\,75332 for Jan-Feb 2007 to Jan 2019. Radial, azimuthal and meridional components are depicted from left to right for each epoch. Ticks along the lower axes indicate the phases of observations. The maps shaded in grey have been reconstructed from limited observations, and this has been considered in our analysis of the ZDI results. Model fits corresponding to each set of maps are shown in Figure \ref{fig:ZDIfits}.}
\label{fig:ZDImaps}
\end{figure*}

\begin{figure*}
   \centering
   \includegraphics[width=0.75\linewidth]{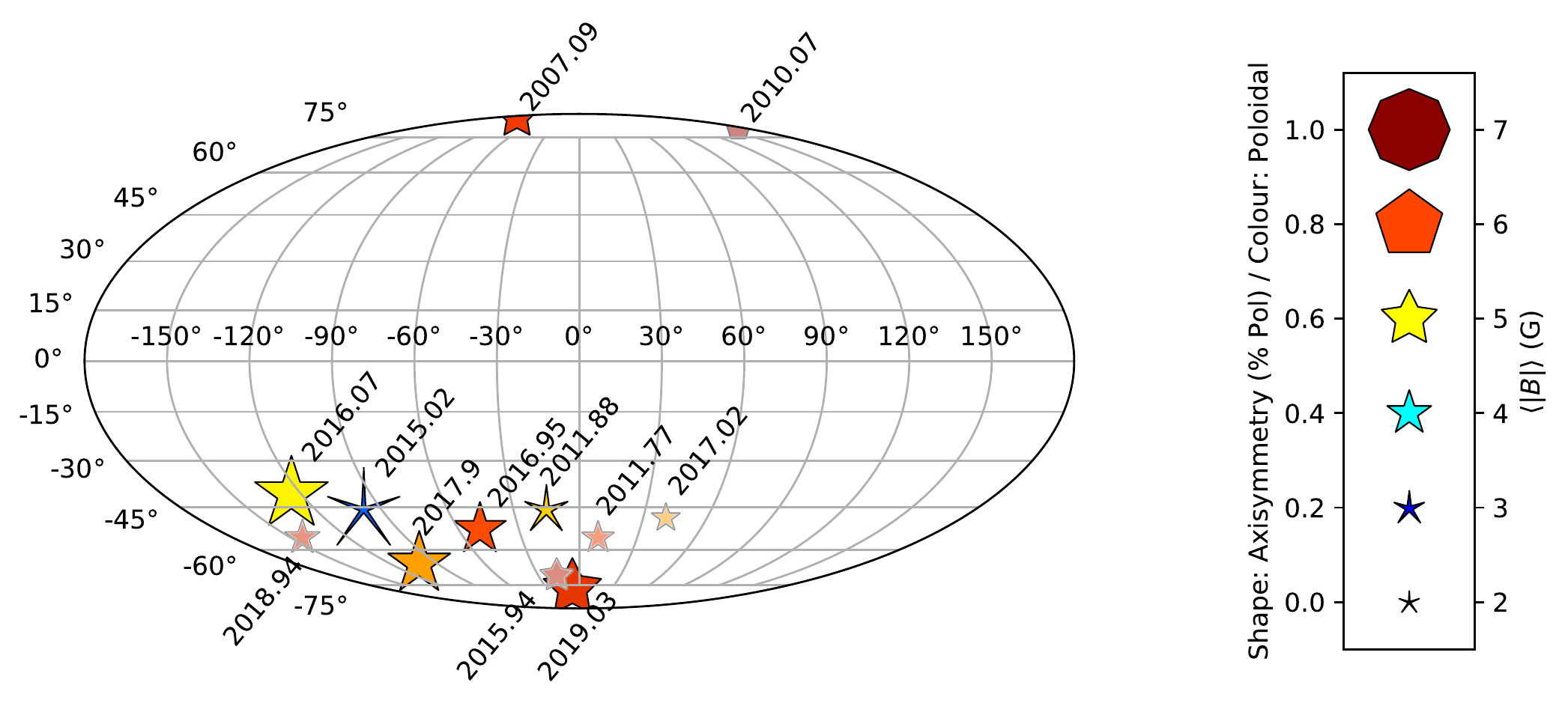}
   \caption{Key geometric properties of the reconstructed large-scale surface magnetic field for each epoch. The positions of each data point correspond to the location of the positive pole of the radial component of the large-scale dipole. The longitudinal position of the dipole should be taken as indicative only; the exact longitude is difficult to determine for a star with significant rotational shear, particularly for epochs where the dipole latitude is low. Symbol size represents the mean unsigned magnetic field strength, $\langle{|B|}\rangle$. Colour and shape represent the fractions of magnetic energy stored in poloidal and axisymmetric poloidal modes (red to blue: poloidal to toroidal, octagon to asterisk: axisymmetric to nonaxisymmetric). For epochs with a limited number of observations (Jan 2010, Oct 2011, Dec 2015, Jan 2017 and Dec 2018) the symbols have a reduced opacity. }
\label{fig:FieldGeom}
\end{figure*}

The reconstructed magnetic field maps are shown in Figure \ref{fig:ZDImaps} and corresponding model fits to the observed LSD profiles are shown in Figure \ref{fig:ZDIfits}. Figure \ref{fig:FieldGeom} shows the location of the positive pole of the large-scale dipole for each epoch, along with the mean unsigned magnetic field strength and fractions of magnetic energy stored in poloidal and axisymmetric modes. Full details of the derived field geometry are provided in Table \ref{tab:fieldGeom}, including the possible variations in the field properties that would result from varying the model parameters ($i\pm\,5\degr$, $v\sin{i}\pm\,0.5\textrm{\,km\,s}^{-1}$, $RV\pm\,0.1\textrm{\,km\,s}^{-1}$, ${P_{rot}}^{+0.11}_{-0.14}\textrm{\,d}$, $d\Omega^{+0.23}_{-0.22}\textrm{\,rad\,d}^{-1}$ and $\chi^2_{aim}\pm\,0.05$, see Section \ref{section:Sensitivity analysis} for further details).

\begin{table*}
\caption{Geometric properties of the large-scale magnetic field of HD\,75332 for Jan-Feb 2007 through to Jan 2019. Columns (left to right) indicate the unsigned mean and maximum surface field strengths, the fraction of the total magnetic energy stored in the poloidal component, the percentages of poloidal and toroidal energy stored in dipolar ($l=1$), quadrupolar ($l=2$) and octopolar modes ($l=3$), the percentages of the total, poloidal and toroidal field energies stored in axisymmetric modes and the location of the positive pole of the radial component of the large-scale dipole. The longitudinal position of the dipole should be taken as indicative only; the exact longitude is difficult to determine for a star with significant rotational shear, particularly for epochs where the dipole latitude is low.  The upper and lower values shown for each measured parameter represent their sensitivity to variations in the model parameters. They were derived by simultaneously varying inclination angle ($\pm\,5\degr$), $v\sin{i}$ ($\pm\,0.5\textrm{\,km\,s}^{-1}$), RV ($\pm\,0.1\textrm{\,km\,s}^{-1}$),  $P_{rot}$ ($^{+0.11}_{-0.14}\textrm{\,d}$), $d\Omega$ ($^{+0.23}_{-0.22}\textrm{\,rad\,d}^{-1}$) and $\chi^2_{aim}$ ($\pm\,0.05$), in consideration of their dependencies, and taking the extremes of all possible values.}
\setlength{\tabcolsep}{5pt}
\begin{tabular}{lcccccccccccccc}
\toprule
\multirow{2}{*}{\textbf{Epoch}} & {\textbf{Bmean}} & {\textbf{Bmax}} & \multicolumn{4}{c}{\textbf{Poloidal}}              & \multicolumn{3}{c}{\textbf{Toroidal}}                        & \multicolumn{3}{c}{\textbf{Axisymmetric}}                             & \multicolumn{2}{c}{\textbf{Dipole radial +}}              \\ \cmidrule(lr){4-7} \cmidrule(lr){8-10} \cmidrule(lr){11-13} \cmidrule(lr){14-15}
                                   &  {\textbf{(G)}} & {\textbf{(G)}}                    & \textbf{\% tot} & \textbf{$l=1$} & \textbf{$l=2$} & \textbf{$l=3$} & \textbf{$l=1$} & \textbf{$l=2$} & \textbf{$l=3$} & \textbf{\% tot} & \textbf{\% Pol} & \textbf{\% Tor} & \textbf{Colat.} & \textbf{Long.} \\ \midrule
                                 
Jan-Feb 2007 	&	3.8$^{+0.3}_{-0.8}$ &16.5$^{+4.3}_{-4.9}$ &83$^{+6}_{-17}$ &55$^{+8}_{-20}$ &25$^{+7}_{-3}$ &13$^{+7}_{-3}$ &43$^{+5}_{-11}$ &39$^{+3}_{-4}$ &13$^{+6}_{-2}$ &79$^{+4}_{-16}$ &79$^{+4}_{-21}$ &80$^{+5}_{-9}$ &4$^{+2}_{-0}$ &41$^{+26}_{-9}$ \\ [+1.5mm]
Jan 2010 		&	2.2$^{+0.2}_{-0.5}$ &9.4$^{+1.6}_{-2.2}$ &97$^{+2}_{-3}$ &72$^{+8}_{-16}$ &18$^{+6}_{-4}$ &7$^{+5}_{-3}$ &36$^{+12}_{-7}$ &48$^{+4}_{-10}$ &11$^{+6}_{-3}$ &85$^{+4}_{-7}$ &86$^{+4}_{-7}$ &42$^{+23}_{-7}$ &10$^{+6}_{-2}$ &352$^{+20}_{-13}$ \\ [+1.5mm]
Oct 2011 		&	2.9$^{+1.0}_{-0.4}$ &8.9$^{+3.3}_{-1.1}$ &83$^{+7}_{-14}$ &67$^{+7}_{-13}$ &22$^{+5}_{-5}$ &9$^{+5}_{-3}$ &52$^{+10}_{-9}$ &37$^{+5}_{-7}$ &9$^{+4}_{-3}$ &55$^{+8}_{-13}$ &51$^{+10}_{-25}$ &78$^{+11}_{-9}$ &146$^{+6}_{-18}$ &190$^{+4}_{-4}$ \\ [+1.5mm]
Nov 2011 		&	4.3$^{+2.7}_{-0.3}$ &14.3$^{+7.1}_{-0.9}$ &65$^{+7}_{-24}$ &48$^{+6}_{-26}$ &20$^{+0}_{-9}$ &12$^{+16}_{-6}$ &55$^{+7}_{-14}$ &32$^{+3}_{-4}$ &10$^{+8}_{-6}$ &40$^{+21}_{-12}$ &24$^{+12}_{-5}$ &70$^{+21}_{-24}$ &136$^{+18}_{-12}$ &165$^{+17}_{-34}$ \\ [+1.5mm]
Jan 2015 		&	7.2$^{+2.8}_{-0.8}$ &18.5$^{+10.0}_{-3.1}$ &28$^{+6}_{-4}$ &24$^{+13}_{-10}$ &17$^{+6}_{-5}$ &26$^{+2}_{-4}$ &61$^{+7}_{-7}$ &31$^{+4}_{-6}$ &5$^{+4}_{-3}$ &71$^{+2}_{-6}$ &17$^{+3}_{-6}$ &92$^{+2}_{-3}$ &136$^{+6}_{-17}$ &82$^{+10}_{-19}$ \\ [+1.5mm]
Dec 2015 		&	3.0$^{+0.7}_{-0.8}$ &14.2$^{+4.3}_{-3.8}$ &92$^{+3}_{-13}$ &61$^{+13}_{-20}$ &22$^{+5}_{-5}$ &10$^{+7}_{-5}$ &29$^{+20}_{-22}$ &40$^{+4}_{-9}$ &20$^{+10}_{-8}$ &75$^{+10}_{-15}$ &77$^{+9}_{-14}$ &60$^{+28}_{-29}$ &161$^{+3}_{-9}$ &164$^{+11}_{-13}$ \\ [+1.5mm]
Jan 2016 		&	6.6$^{+2.1}_{-0.8}$ &18.2$^{+5.8}_{-2.2}$ &61$^{+6}_{-12}$ &46$^{+10}_{-4}$ &24$^{+3}_{-6}$ &16$^{+2}_{-3}$ &62$^{+6}_{-9}$ &33$^{+3}_{-5}$ &5$^{+4}_{-3}$ &72$^{+4}_{-2}$ &57$^{+6}_{-7}$ &95$^{+2}_{-3}$ &130$^{+16}_{-11}$ &56$^{+5}_{-5}$ \\ [+1.5mm]
Dec 2016 		&	4.7$^{+2.1}_{-0.7}$ &16.9$^{+12.3}_{-3.3}$ &79$^{+6}_{-17}$ &48$^{+10}_{-30}$ &17$^{+5}_{-8}$ &13$^{+6}_{-2}$ &41$^{+13}_{-24}$ &31$^{+3}_{-8}$ &17$^{+13}_{-7}$ &47$^{+12}_{-25}$ &44$^{+13}_{-29}$ &59$^{+13}_{-27}$ &142$^{+14}_{-18}$ &131$^{+11}_{-10}$ \\ [+1.5mm]
Jan 2017 		&	2.5$^{+0.5}_{-0.4}$ &9.5$^{+1.2}_{-1.2}$ &71$^{+7}_{-9}$ &66$^{+4}_{-9}$ &21$^{+3}_{-2}$ &8$^{+3}_{-1}$ &64$^{+3}_{-5}$ &29$^{+2}_{-1}$ &6$^{+3}_{-1}$ &54$^{+6}_{-10}$ &45$^{+7}_{-13}$ &78$^{+6}_{-9}$ &139$^{+3}_{-9}$ &221$^{+7}_{-5}$ \\ [+1.5mm]
Nov-Dec 2017 	&	5.7$^{+3.8}_{-1.3}$ &19.8$^{+11.8}_{-2.3}$ &70$^{+5}_{-17}$ &24$^{+5}_{-7}$ &29$^{+5}_{-11}$ &23$^{+5}_{-6}$ &61$^{+6}_{-11}$ &30$^{+6}_{-12}$ &4$^{+7}_{-3}$ &65$^{+6}_{-7}$ &56$^{+8}_{-17}$ &87$^{+7}_{-10}$ &156$^{+2}_{-50}$ &79$^{+50}_{-22}$ \\ [+1.5mm]
Dec 2018 		&	3.1$^{+0.7}_{-0.8}$ &13.6$^{+6.4}_{-2.0}$ &88$^{+6}_{-6}$ &45$^{+10}_{-17}$ &23$^{+4}_{-5}$ &17$^{+5}_{-5}$ &23$^{+16}_{-20}$ &16$^{+11}_{-6}$ &36$^{+6}_{-16}$ &52$^{+10}_{-12}$ &56$^{+9}_{-15}$ &23$^{+24}_{-1}$ &146$^{+6}_{-12}$ &36$^{+17}_{-13}$ \\ [+1.5mm]
Jan 2019 		&	5.2$^{+2.8}_{-0.9}$ &21.1$^{+9.4}_{-4.4}$ &84$^{+6}_{-21}$ &58$^{+7}_{-25}$ &21$^{+5}_{-9}$ &11$^{+8}_{-4}$ &57$^{+13}_{-18}$ &28$^{+9}_{-12}$ &9$^{+9}_{-2}$ &69$^{+4}_{-33}$ &74$^{+3}_{-41}$ &41$^{+19}_{-14}$ &166$^{+2}_{-22}$ &174$^{+59}_{-23}$ \\ [+1.5mm]\bottomrule

    \end{tabular}
    \label{tab:fieldGeom}
\end{table*}

The results show that HD\,75332 has a weak large-scale magnetic field with an unsigned average strength that varies between $2.2\,^{+0.2}_{-0.5}$\,G and $7.2\,^{+2.8}_{-0.8}$\,G across our observational epochs. Localized field strengths reach up to $21.1\,^{+9.4}_{-4.4}$\,G.  For most epochs, the radial field dominates over both the azimuthal and meridional components, but HD\,75332 is also capable of producing a strong azimuthal field which dominates in Jan 2015 and Jan 2016. 

\subsubsection{Radial field component}
A key result shown by the magnetic field maps is the reversal of the large-scale radial field polarity between Jan 2010 and Oct 2011. In Jan-Feb 2007 and Jan 2010 a positive radial field dominates over the visible northern pole of HD\,75332 and down to low latitudes. From Oct 2011 onward the radial field has a dominantly negative polarity in the northern hemisphere and positive polarity in the southern hemisphere. The polarity reversal shown in the maps also coincides with the relocation of the positive pole of the radial component of the large-scale dipole across the equator from $80\degr\,^{+2}_{-6}$\,N to $56\degr\,^{+6}_{-18}$\,S (Figure \ref{fig:FieldGeom}), and the reversal of the Stokes {\it{V}} polarization signatures (Figure \ref{fig:ZDIfits}). 

Prior to the observed polarity reversal the radial magnetic field is weak but it dominates over the azimuthal and meridional components, reflecting the large fraction of magnetic energy stored in the poloidal field. The field geometry is simple and strongly axisymmetric in both Jan-Feb 2007 and Jan 2010, and the positive pole of the large-scale radial dipole is located far from the equator.

Following the polarity reversal the radial field varies significantly across epochs in terms of both its structure and strength. The visible hemisphere is dominated by a negative field but the maps also show the emergence of positive polarity regions at the pole in Jan 2015 and Jan 2016, and close to the pole in Nov 2011. At these epochs the positive pole of the radial component of the dipolar axis is located close to the equator, and the azimuthal field is also strong, reflecting a large fraction of magnetic field stored in the toroidal component.  It should also be noted that the high-latitude positive pockets of field are co-located in the radial and meridional maps, which is a possible indicator that they may be artifacts. However, these features appear to be robust in our model sensitivity analysis (Section \ref{section:Sensitivity analysis}) so we are confident that they are real.

The radial field is at its strongest in Dec 2016, Nov-Dec 2017 and Jan 2019, but this may be partly related to the higher number of observations used to reconstruct these maps; the magnitude and complexity of the field recovered from ZDI depends on sufficient sampling of the stellar surface.  Conversely, the limited numbers and poor phase coverage of observations in Jan 2010, Oct 2011, Dec 2015, Jan 2017 and Dec 2018 likely mean that our reconstructed field maps underestimate the strength and complexity of the magnetic field for these epochs. We tested reduced data sets to examine the impacts of differences in data quantity on the reconstructed magnetic field and the apparent evolution of magnetic field properties from epoch to epoch (see Section \ref{subsec:pairs_of_epochs}) .

In addition to the magnetic maps shown, our spectropolarimetric data also contained a number of small data sets with insufficient observations to reliably reconstruct the magnetic field using ZDI. The observed LSD line profiles for these small data sets are included in Figure \ref{fig:ZDIfits2} and provide some insight into the large-scale polarity of the magnetic field. The Stokes {\it{V}} profiles for Nov 2006 are particularly interesting; the polarization signature is opposite that observed in Jan-Feb 2007, suggesting that the large-scale polarity might also be opposite. This may indicate that an additional reversal of the radial field has occurred between Nov 2006 and Jan-Feb 2007, though this cannot be confirmed from our current data. 

\subsubsection{Azimuthal field component}\label{subsubsec:azimuthal}

At most epochs the azimuthal component is dominated by a single polarity and has a strong axisymmetric component. Multiple reversals of the azimuthal field occur across our $\sim$13 years of observations. The azimuthal field is weakly negative in Jan-Feb 2007 and Jan 2010. Between Oct 2011 and Jan 2017, the maps show a positive azimuthal field with varying strength. Then, in Nov-Dec 2017 the azimuthal field is negative, followed by a mixed polarity in Dec 2018 and then a predominantly positive polarity in Jan 2019. Although both the azimuthal and radial field components reverse between Jan 2010 and Oct 2011, there is no systematic correlation between the polarity of the large-scale radial dipole and the sign of the azimuthal field. A strengthening of the azimuthal field appears to coincide with the emergence of the high-latitude positive radial regions in Nov 2011, Jan 2015 and Jan 2016. 

\subsubsection{Meridional field component}

There appears to be significant cross-talk between the radial and meridional components of the magnetic field, with many of the smaller-scale features of the radial field reproduced in the meridional field maps. This is to be expected for stars with $i\,\leq\,30\degr$ \citep{Donati1997b}. It is unlikely that this cross-talk would impact on the major findings of this study, which are based on the recovered large-scale magnetic field structure rather than smaller-scale magnetic features. 

\subsection{Effects of model parameters on the reconstructed magnetic field}\label{section:Sensitivity analysis}

\begin{figure}
\centering
    \includegraphics[width=\columnwidth]{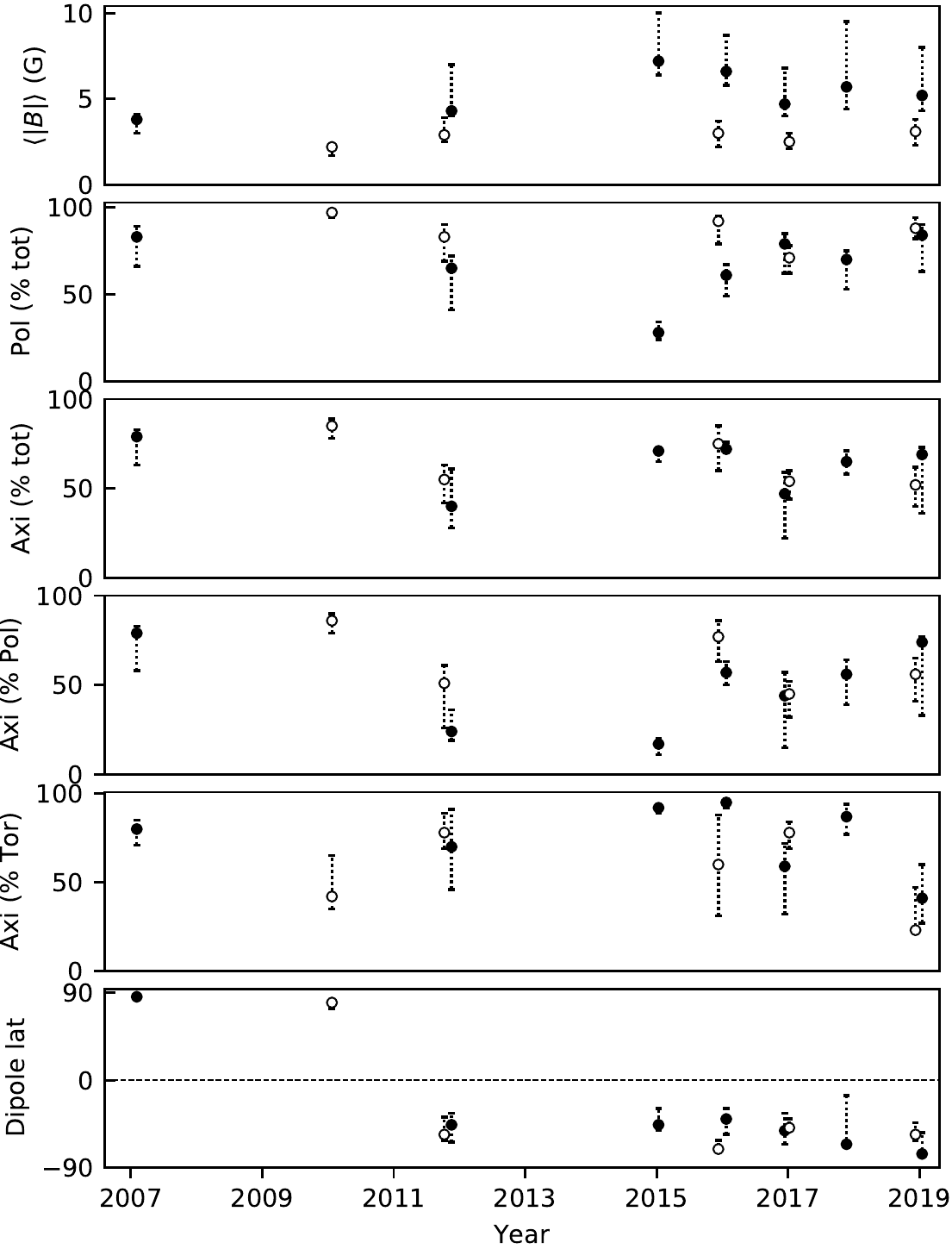}
    \caption{Time-series variability of the key magnetic field properties shown in Figure \ref{fig:FieldGeom}. Additionally, the total axisymmetric field and toroidal axisymmetric field strengths across epochs are shown. Open circles are used for epochs where limited observations were available for ZDI. The dotted bars shown are not formal error bars; rather, they indicate the sensitivity of the derived magnetic field properties to variations in the model input parameters. They were determined by simultaneously varying the model parameters within the following limits; inclination angle ($\pm\,5\degr$), $v\sin{i}$ ($\pm\,0.5\textrm{\,km\,s}^{-1}$), RV ($\pm\,0.1\textrm{\,km\,s}^{-1}$), $P_{rot}$ ($^{+0.11}_{-0.14}\textrm{\,d}$), $d\Omega$ ($^{+0.23}_{-0.22}\textrm{\,rad\,d}^{-1}$) and $\chi^2_{aim}$ ($\pm\,0.05$), and taking into account the dependency between $P_{rot}$ and $d\Omega$. Note that the bars do not indicate the sensitivity of epoch-to-epoch trends to variations in the model parameters.} 
    \label{fig:FieldGeomUncertainty}
\end{figure}

We tested the sensitivity of our ZDI results to changes in the adopted stellar parameters by simultaneously varying the parameters within their derived uncertainty domains. We tested $i\pm\,5\degr$, $v\sin{i}\pm\,0.5\textrm{\,km\,s}^{-1}$ and $RV\pm\,0.1\textrm{\,km\,s}^{-1}$. To account for the dependency between $P_{rot}$ and $d\Omega$, we tested four combinations of the DR parameters that correspond to the extremes of the 1-$\sigma$ paraboloid shown in Figure \ref{fig:DR_2015} (left). We consider this to be a conservative range given that the S-index variability suggests a minimum $d\Omega$ of 0.15\,rad\,d$^{-1}$ (Section \ref{sec:diffrot}).  Finally, we varied the target $\chi^2$ value ($\chi^2_{aim}\pm0.05$), with this range carefully selected to ensure that all models could converge to the specified $\chi^2_{aim}$ and avoid over-fitting to noise in the Stokes {\it{V}} profiles. 

\begin{figure*}
\includegraphics[width=\linewidth]{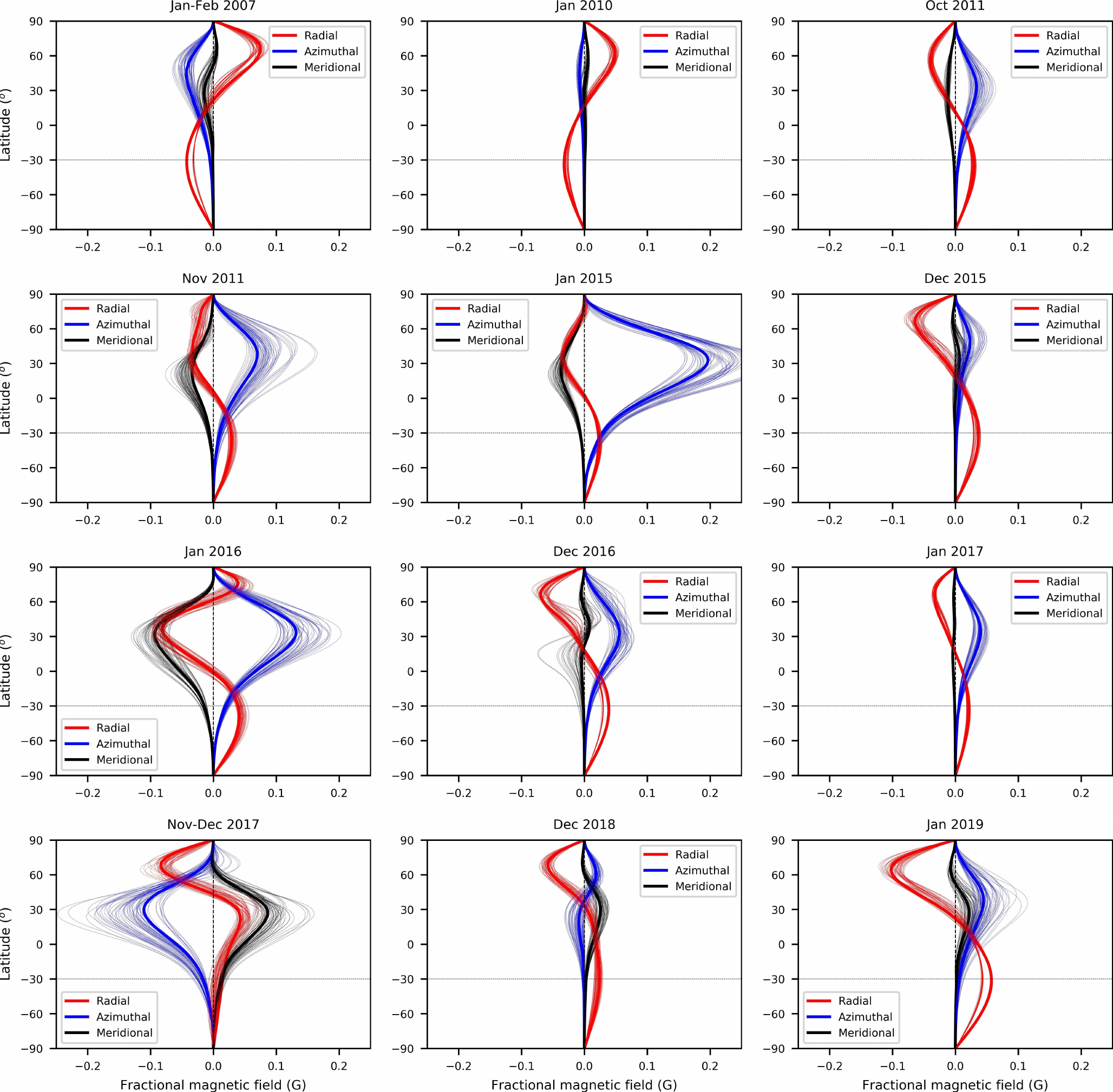}
\caption{Latitude vs fractional magnetic energy stored in radial (red), azimuthal (blue) and meridional (black) fields for Jan-Feb 2007 to Jan 2019. Thick lines represent the result using the optimum stellar and model parameters. Thin lines represent the results for simultaneous variations in inclination angle ($\pm\,5\degr$), $v\sin{i}$ ($\pm\,0.5\textrm{\,km\,s}^{-1}$), RV ($\pm\,0.1\textrm{\,km\,s}^{-1}$), $P_{rot}$ ($^{+0.11}_{-0.14}\textrm{\,d}$), $d\Omega$ ($^{+0.23}_{-0.22}\textrm{\,rad\,d}^{-1}$) and $\chi^2_{aim}$ ($\pm\,0.05$), taking into account the dependency between $P_{rot}$ and $d\Omega$. The dashed lines at $-30\degr$ latitude correspond to the lower extents of our ZDI maps in Figure \ref{fig:ZDImaps}.}
\label{fig:FractionalMagField1}
\end{figure*}

The impacts of the model parameter variations on the magnetic field properties from Figure \ref{fig:FieldGeom} are shown in Figure \ref{fig:FieldGeomUncertainty}. It is important to note that the dotted `variation bars' shown in Figure \ref{fig:FieldGeomUncertainty} are not formal error bars, and that the robustness of epoch-to-epoch changes in the magnetic field cannot be directly inferred by comparing the bars for consecutive epochs. Additionally, Figure \ref{fig:FractionalMagField1} shows the latitudinal distributions of the magnetic field for each of the radial, azimuthal and meridional components, and their sensitivity to variations in the model parameters. 
We calculated field strength as a function of latitude for each field component using Equation \ref{eq:frac_mag_energy} \citep{Waite2015}.
\begin{equation}\label{eq:frac_mag_energy}
    F(\theta)=\frac{B(\theta)\cos{(\theta)}d\theta}{2}
\end{equation}
where $F(\theta)$ and $B(\theta)$ are the fractional and average magnetic field strengths of each component at latitude $\theta$, and $d\theta$ is the thickness of individual latitude bands.

The strength and complexity of the reconstructed magnetic field are most sensitive to changes in the $\chi^2_{aim}$. Adopting a low $\chi^2_{aim}$ of 0.90 improves the fit of the model to small features in the Stokes {\it{V}} line profiles, but we can not rule out that these features may be related to noise.  Sensitivity tests using $\chi^2_{aim}=0.90$ are the predominant cause of the increased mean field strength and decreased poloidal field strength indicated by the variation bars in Figure \ref{fig:FieldGeomUncertainty}. This effect is also reflected in Figure \ref{fig:FractionalMagField1} as an increased azimuthal field strength across all latitudes.

Importantly, our tests show that the polarity of the large-scale magnetic field recovered from ZDI is not sensitive to variations in the stellar and model parameters within the ranges we tested. Figure\,\ref{fig:FieldGeomUncertainty} shows that even when we vary the model parameters within a large domain we still observe the reversal of the large-scale radial dipole between Jan 2010 and Oct 2011. Likewise, the magnetic energy distributions in Figure \ref{fig:FractionalMagField1} show a switch from a dominantly positive radial field polarity in the northern hemisphere to a negative polarity between Jan 2010 and Oct 2011 for all tested models. The multiple reversals of the azimuthal field component we described in Section \ref{subsubsec:azimuthal} are also shown in Figure \ref{fig:FractionalMagField1} for all variations of the model parameters. 

Other key results in the ZDI maps were the emerging high-latitude positive regions in the radial field in Nov 2011, Jan 2015 and Jan 2016. When varying the model parameters, the magnetic maps for Nov 2011 consistently show high-latitude positive radial regions of varying strength. However, it appears that the positive regions may be dominated by negative polarity regions at the same latitudes, such that Figure \ref{fig:FractionalMagField1} shows a negative average radial field at high latitudes for all of the tested models. For Jan 2015 and Jan 2016, dominant positive polarity regions are reconstructed at high latitudes in the radial field even with significant variations in the model parameters, as is shown by Figure \ref{fig:FractionalMagField1}.  Since these features show a low sensitivity to variations in the model input parameters, we are confident that they are real.  In Section \ref{subsec:pairs_of_epochs} we further discuss how variations in the stellar and model parameters, and differences in the number of observations used for each ZDI map, impact on the observed evolution of the magnetic field from epoch to epoch.

\subsection{Longitudinal magnetic field}

The longitudinal magnetic field, $B_l$ (G), is the line-of-sight field strength averaged over the visible stellar surface. For an inclined star, the $B_l$ will be dominated by the average of the field on the visible hemisphere. Therefore, if the field is mainly axisymmetric and poloidal, $B_l$ can be used to trace magnetic polarity reversals throughout the magnetic cycle. Though, if the poloidal field is non-axisymmetric (e.g. during Jan 2015 for HD\,75332), the $B_l$ would also be modulated by the phase of rotation that changes the visibility of non-axisymmetric components. 

\begin{figure}
    \centering
    \includegraphics{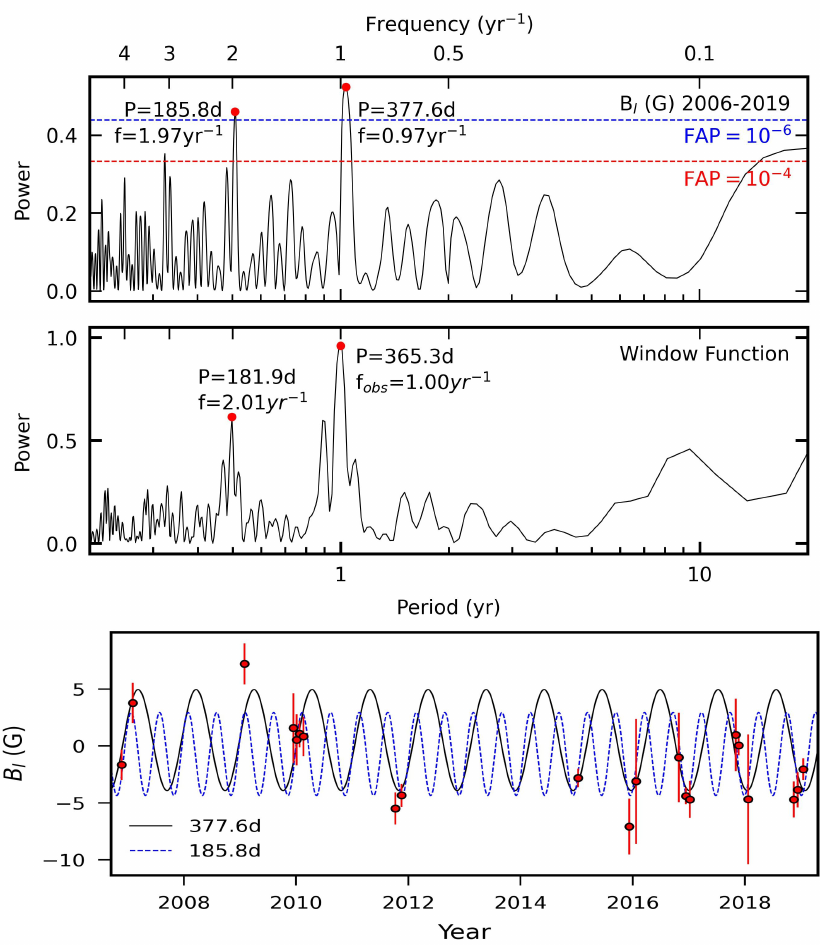}
    \caption{Top: Power spectrum for the mean longitudinal magnetic field, $B_l$ (G), computed using the GLS periodogram \citep{Zechmeister2009}. 
    Second: Power spectrum for the Window Function.
    Bottom: Time series of $B_l$ observations showing the fit of the 377.6\,d (black line) and 185.8\,d (blue dashed line) periodicities to the observations. For clarity, we show the mean $B_l$ and its propagated error for each observational epoch listed in Table \ref{tab:NARVALdata} rather than the individual observations (Table \ref{tab:NARVALobsdetails}) that were used to compute the periodogram.}
    \label{fig:Bl}
\end{figure}

We measured $B_l$ directly from the NARVAL Stokes {\it{V}} and {\it{I}} LSD profiles using Equation \ref{eq:blong} from \citet{Donati1997}.
\begin{equation}\label{eq:blong}
    B_{l}=-2.14\times{10^{11}} \frac{\int{vV(v) dv}}{\lambda g c \int{[1-I(v)]dv}}
\end{equation}
where {\it{V(v)}} and {\it{I(v)}} are the Stokes {\it{V}} and {\it{I}} LSD profiles respectively, $\lambda$ (570\,nm) is the central wavelength of the LSD profile, $g$ is the mean Land\'e factor (1.21), c is the speed of light in km\,s$^{-1}$ and $v$ is velocity in km\,s$^{-1}$. We integrated over a velocity domain of -18 to 27\,km\,s$^{-1}$ to include the entire Stokes {\it{V}} polarization signal while minimizing noise. The derived $B_l$ for each observation is shown in Table \ref{tab:NARVALobsdetails} with error bars determined by propagating the uncertainties computed during the reduction process for each spectral bin of the normalized spectrum through Equation \ref{eq:blong}. Our values range from $-9.3\pm2.8$\,G to $8.2\pm1.8$\,G, excluding observations with very large error bars owing to poor SNRs. Mean $B_l$ values for each epoch in Table \ref{tab:NARVALdata} are shown as a time-series in Figure \ref{fig:Bl}. \citet{marsden2014} used a similar method to determine $B_l$ for the NARVAL data we present here for Nov 2006 to Nov 2011, measuring a maximum  ${|B_l|=8.1\pm2.6}$\,G. 

Comparing the measured $B_l$ to the magnetic maps, the sign of $B_l$ is generally consistent with the polarity of the large-scale radial magnetic field over the visible northern pole of HD\,75332. For Nov 2006, where we had insufficient observations to reconstruct a map of the magnetic field, the measured $B_l$ supports a negative radial field polarity over the visible pole. This strengthens the case for a reversal of the radial field component between Nov 2006 and Jan-Feb 2007, in addition to the reversal shown in our ZDI results between Jan 2010 and Oct 2011. 

We also carried out a period search for the $B_l$ values using the same process described in Section \ref{subsec:S-index period search}. We applied the GLS periodogram to all $B_l$ observations in Table \ref{tab:NARVALobsdetails} and the resulting power spectrum is shown in Figure \ref{fig:Bl}. The FAPs indicated in Figure \ref{fig:Bl} were determined as described in Section \ref{subsec:S-index period search} using $10^{6}$ re-sampled data sets (measurement errors were preserved during re-sampling). Figure \ref{fig:Bl} also shows the power spectrum computed for the Window Function ($B_l=1$), which shows a significant annual frequency (hereafter $f_{obs}$) related to the seasonal observations, and a 6-month frequency that is likely an alias of $f_{obs}$. We detected multiple significant periodicities in the $B_l$ data, separated by frequency intervals precisely equal to $f_{obs}$, which suggests that they are a series of aliases of one `true' $B_l$ frequency \citep{VanderPlas2018}. The two most significant peaks correspond to signals with $P=185.8\pm0.6$\,d ($\textrm{amplitude}=3.7\pm0.4$\,G) and $P=377.6\pm3.9$\,d ($\textrm{amplitude}=4.4\pm0.4$\,G). We compare both periodicities to the time-series data in Figure \ref{fig:Bl}, but it is not clear which provides a better fit to the observations. Whichever is the `true' signal, the low FAP suggests that it is distinct from the annual peak in the Window Function.  There is also scatter in the time series data from epoch to epoch, which suggests that additional longer periodicities may be present, though these do not clearly correspond to any peaks in the power spectrum. 

\section[Magnetic field vs chromospheric activity]{Magnetic field \MakeLowercase{vs} chromospheric activity}\label{sec:Magnetic field vs chromospheric activity]}

\subsection{$B_l$ vs S-index variability}
The chromospheric S-index and mean longitudinal magnetic field strength ($B_l$) of HD\,75332 appear to vary cyclically and on possibly proportional time scales. We detected a $\sim$193.5\,d chromospheric activity cycle in the S-index. For the $B_l$ data we detected a series of related periodicities, with the highest power signal corresponding to a period of 377.6\,d, roughly double (within 3\%) the length of the S-index cycle. We also detected 3.9\,yr and 31.5\,yr cycles in the chromospheric S-index, but it is not clear from our data if these might correspond to $B_l$ cycles.

\subsection{Large-scale field polarity vs S-index variability}
\begin{figure*}
    \centering
    \includegraphics{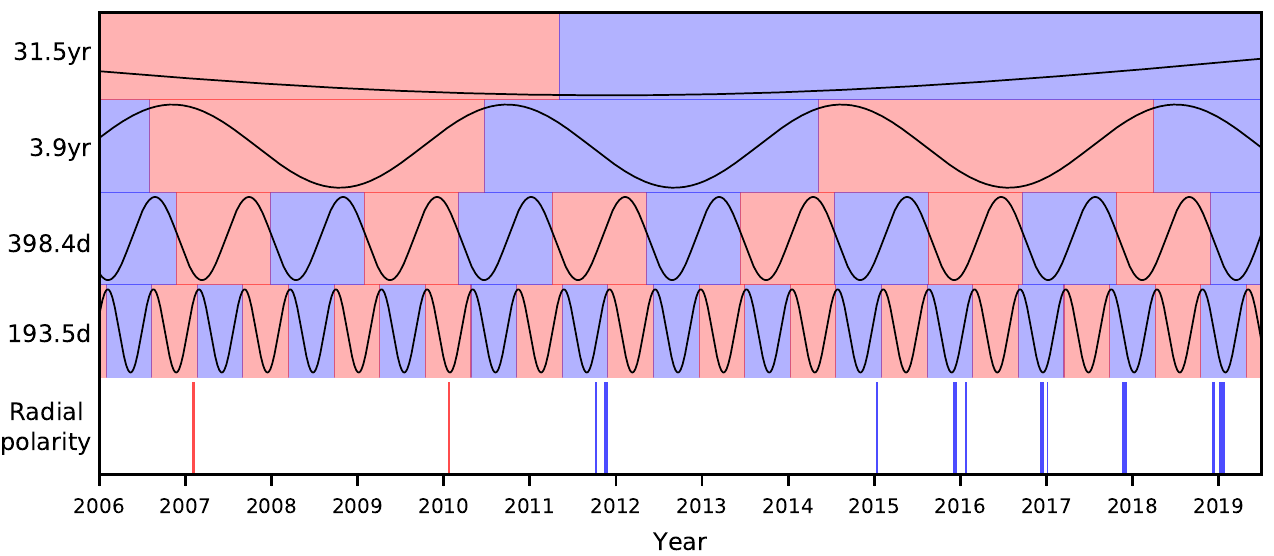}
    \caption{Observed radial field polarity (bottom) versus expected polarity for magnetic half-cycles of 31.5\,yr (top), 3.9\,yr (second row), 398.4\,d (third row) and 193.5\,d (fourth row). Each cycle corresponds to a significant peak in our S-index power spectrum (Figure \ref{fig:Sindex}). The frequency and horizontal shift of each cycle were fitted to the S-indices using the GLS periodogram code-base by \citet{Zechmeister2009}. The shaded regions represent opposite polarities of the radial magnetic field. Note that, in reality, activity and magnetic cycles are not strictly periodic and may not maintain their periods over the time-frame of our observations. For the observed polarity, the shading spans the dates of observations used for each magnetic field reconstruction.}
    \label{fig:Polarity}
\end{figure*}

Our ZDI results allow us to probe the relationship between the magnetic field geometry and chromospheric activity cycles. Figure \ref{fig:Polarity} compares the large-scale radial field polarity in our reconstructed maps to the S-index cycles. Each detected S-index cycle is modelled by a periodic function which was fitted to the S-index observations using the GLS periodogram (Section \ref{subsec:S-index period search}). Note that the expected magnetic field polarity shown in Figure \ref{fig:Polarity} is based on the assumption that the large-scale field polarity oscillates with the S-index cycle. The cycles are assumed to maintain their periods and amplitudes over the time-frame of our observations, though in reality activity and magnetic cycles are not strictly periodic and may not keep their phase coherence over this time-scale. Figure \ref{fig:Polarity} shows that the observed radial field polarity is consistent with reversals occurring in phase with the 31.5\,yr S-index cycle, or the 193.5\,d S-index cycle. The observations do not exclude the possibility of a rapid magnetic cycle, despite the fact that we observed only one reversal of the large-scale radial field component. Figure \ref{fig:Polarity} also shows that the observed field polarity is not consistent with reversals occurring in phase with the 398.4\,d or 3.9\,yr S-index cycles.

If we assume that a 31.5\,yr magnetic half-cycle is present in HD\,75332 then the radial polarity reversal we observed between Jan 2010 and Oct 2011 would be the only one to have occurred during our $\sim$13 years of observations. Based on the modelled 31.5\,yr S-index cycle shown in Figure \ref{fig:Polarity}, the polarity reversal roughly coincided with the 31.5\,yr S-index cycle minimum. This would be in contrast to the solar case, where reversals of the large-scale radial field polarity occur at around chromospheric activity cycle maxima. If a 31.5\,yr magnetic half-cycle is present for HD\,75332, then the mean field strength appears to roughly correlate with the S-index, as it does in the Sun. The field strength is lower in Jan-Feb 2007 and Jan 2010 prior to activity cycle minimum, and generally stronger following (Figure \ref{fig:FieldGeomUncertainty}). However, the structure of the magnetic field would appear to evolve differently throughout cycles compared to the solar case. We do not observe a solar-like transition from a complex, high-order multi-polar field prior to a polarity reversal, to a simple, axisymmetric and predominately dipolar field following a reversal.  Rather, the large-scale field of HD\,75332 is simple before the polarity reversal, with a large fraction of the magnetic energy stored in the poloidal, dipolar field. Following the polarity reversal, the field geometry undulates between complex and simple. For example, the poloidal component of the field varies rapidly from $28\,^{+6}_{-4}$\,\% in Jan 2015 to $92\,^{+3}_{-13}$\,\% in Dec 2015.  
 
 We consider it to be much more likely that the large-scale radial field polarity varies in phase with the $\sim$193.5\,d S-index cycle, which would indicate a total magnetic cycle of $\sim$1.06\,yr (equal to two 193.5\,d activity cycles). This would be consistent with the 377.6\,d ($\sim$1.03\,yr) cycle we detected in the $B_l$. If this rapid magnetic cycle is present, Figure \ref{fig:Polarity} suggests that reversals of the radial field may occur at around S-index cycle maxima, similar to the Sun, $\tau$\,Boo and 61 Cyg A.  
 
\begin{figure}
\centering
\includegraphics[width=\columnwidth]{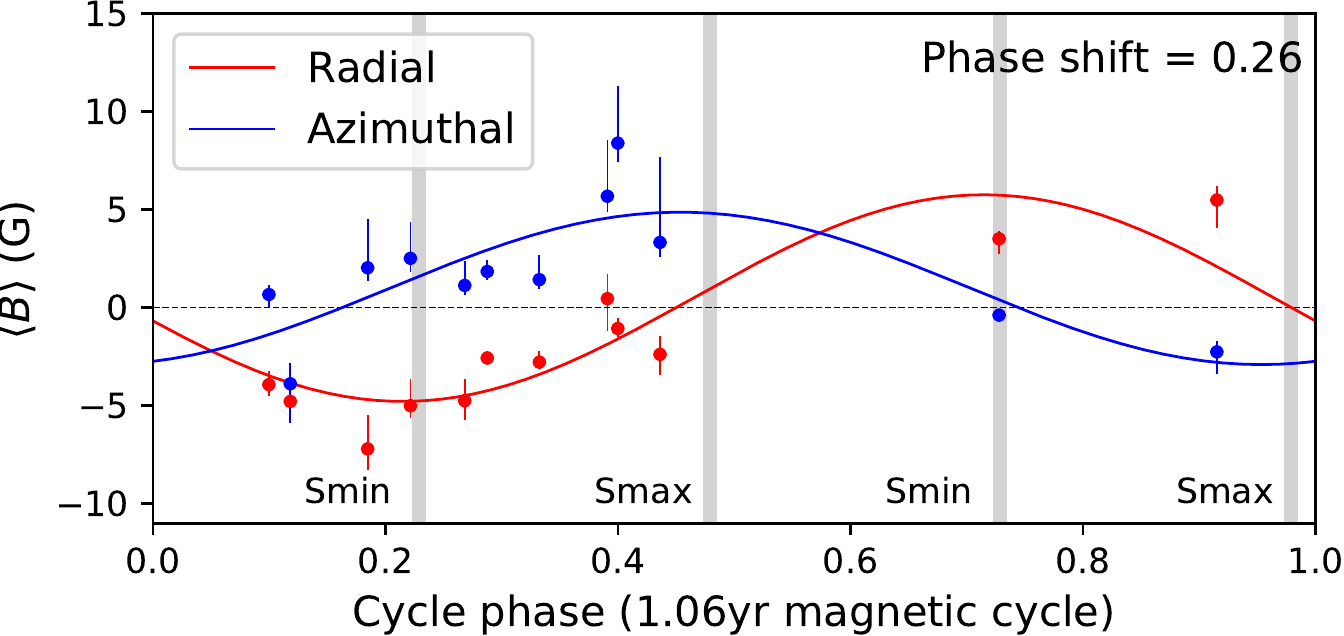}
\caption{Radial (red) and azimuthal (blue) field strengths, averaged over the Northern hemisphere of HD\,75332, throughout a 1.06\,yr magnetic cycle. Variation bars were determined by simultaneously varying the model parameters within our derived uncertainty limits; inclination angle ($\pm\,5\degr$), $v\sin{i}$ ($\pm\,0.5\textrm{\,km\,s}^{-1}$), RV ($\pm\,0.1\textrm{\,km\,s}^{-1}$), $P_{rot}$ ($^{+0.11}_{-0.14}\textrm{\,d}$), $d\Omega$ ($^{+0.23}_{-0.22}\textrm{\,rad\,d}^{-1}$) and $\chi^2_{aim}$ ($\pm\,0.05$). Grey lines represent S-index maxima and minima within the 193.5\,d activity cycle.}
\label{fig:PolarityAzi}
\end{figure}

 In addition to reversals of the radial field polarity, our results show multiple reversals of the large-scale azimuthal field component. The azimuthal field reversal that occurred between Jan 2017 and Nov-Dec 2017, and another possibly between Dec 2018 and Jan 2019, were observed without the concurrent reversal of the radial field, suggesting that the azimuthal and radial flux cycles may not be in phase with one another. We investigate this possibility further in Figure \ref{fig:PolarityAzi}, which compares the average radial and azimuthal field intensities over the visible hemisphere of HD\,75332 for each of our ZDI epochs. The observations have been phase-folded for a 1.06\,yr magnetic cycle and each data set is fitted with a sine function with a period of 1.06\,yr. There is significant scatter of the data about the fitted sine waves due to the fact that our observations may cover as many as 12 magnetic cycles, and additional, longer-term periodicities are likely present in the data. The observed polarity of the azimuthal field is compatible with a 1.06\,yr azimuthal flux cycle that leads the radial flux cycle by around 26\,\% of a magnetic cycle, or 0.28\,yrs. This may be similar to the 25\,\% phase shift present in the Sun between toroidal and poloidal flux cycles \citep{Charbonneau1997,Jouve2007}.  \citet{Jeffers2018} also speculated that there may be a similar lag between large-scale azimuthal and radial field reversals for $\tau$\,Boo. \citet{Fares2009} observed the concurrent reversal of all three components of the magnetic field of $\tau$\,Boo, though they also found toroidal flux cycles to be phase shifted with respect to poloidal flux cycles by 18\,\% using a similar process as we used for Figure \ref{fig:PolarityAzi}. 

\subsection{Epoch-to-epoch magnetic field variability}\label{subsec:pairs_of_epochs}

\begin{figure*}
    \centering
    \includegraphics[width=\linewidth]{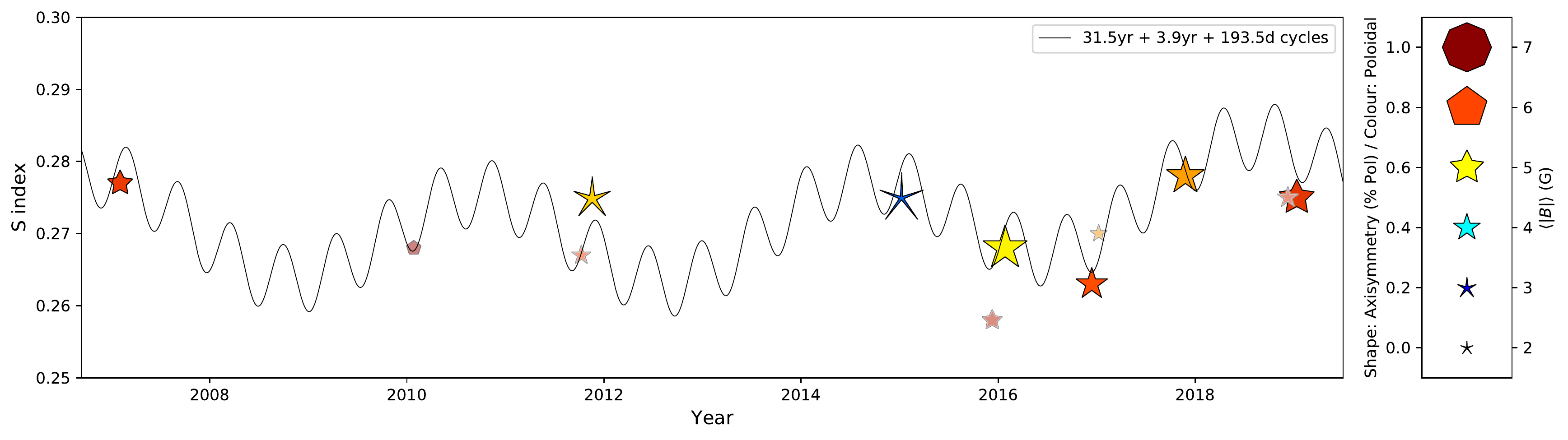}
    \caption{Variations in key magnetic field properties throughout S-index cycles. The black line is a combination of the 193.5\,d, 3.9\,yr and 31.5\,yr periodic functions that were fitted to the S-index observations using the GLS periodogram (Section \ref{subsec:S-index period search}).The size of each symbol represents the mean magnetic field strength. Colour and shape represent the fractions of magnetic energy stored in poloidal and axisymmetric poloidal modes (red to blue: poloidal to toroidal, octagon to asterisk: axisymmetric to nonaxisymmetric). The position of each data point corresponds to the mean S-index for that epoch. For epochs with a limited number of observations (Jan 2010, Oct 2011, Dec 2015, Jan 2017 and Dec 2018) the symbols have a reduced opacity.}
    \label{fig:FieldGeomSindex}
\end{figure*}

The case for a rapid magnetic cycle in HD\,75332 is further strengthened by the rapid evolution we observed in the strength, structure and complexity of the magnetic field from epoch to epoch. Figure \ref{fig:FieldGeomSindex} shows the magnetic field properties from Figure \ref{fig:FieldGeom} against our detected S-index cycles. The location of each data point in Figure \ref{fig:FieldGeomSindex} corresponds to the mean S-index for that epoch. We have four sets of observations that may have been taken between a pair of polarity reversals; Oct 2011 \& Nov 2011, Dec 2015 \& Jan 2016, Dec 2016 \& Jan 2017, and Dec 2018 \& Jan 2019. 

\subsubsection{Oct - Nov 2011}

Between Oct 2011 and Nov 2011, Figure \ref{fig:FieldGeomSindex} shows a trend of increasing magnetic field strength and S-index, coinciding with the progression from activity cycle minimum to maximum. Figure \ref{fig:Bl} also shows a possible increase in the mean unsigned $B_l$ between these epochs. The poloidal component of the field decreases from $83\,^{+7}_{-14}\,\%$ in Oct to $65\,^{+7}_{-24}\,\%$ in Nov, as more magnetic energy is stored in the toroidal field, and the axisymmetry of the poloidal component is reduced from $51\,^{+10}_{-25}\,\%$ to $24\,^{+12}_{-5}\,\%$. These trends are robust, as they prevail when varying the model parameters as described in Section \ref{section:Sensitivity analysis}. Additionally, we investigated how differences in the number and phase coverage of observations across epochs may impact on the observed field evolution. We tested a reduced data set for Nov 2011 to provide an equal number of observations in Oct and Nov, sampling the stellar surface over similar time-frames. We found that this did not impact on the nature of the observed increase in the mean field strength, and decrease in the strength and axisymmetry of the poloidal field between Oct and Nov. The axisymmetry of the toroidal field may decrease between Oct and Nov ($78\,^{+11}_{-9}\,\%$ to $70\,^{+21}_{-24}\,\%$), although the nature of this trend is sensitive to changes in the model parameters and differences in data quantity between epochs. Also, Figure \ref{fig:FieldGeom} shows that the radial dipolar axis may move toward the equator between Oct and Nov, coinciding with a decrease in the latitude of peak radial field intensity toward the equator in Figure \ref{fig:FractionalMagField1}. However, this trend is not consistent across all of our sensitivity tests.

\subsubsection{Dec 2015 - Jan 2016}

The evolution of the magnetic field between Dec 2015 and Jan 2016 is similar to that observed between Oct and Nov 2011, again coinciding with a progression from activity minimum toward maximum in Figure \ref{fig:FieldGeomSindex}. Between Dec 2015 and Jan 2016 the mean field strength, S-index and $B_l$ increase simultaneously, and the poloidal energy fraction decreases significantly from $92\,^{+3}_{-13}\,\%$ to $61\,^{+6}_{-12}\,\%$. The poloidal axisymmetry decreases ($77\,^{+9}_{-14}\,\%$ to $57\,^{+6}_{-7}\,\%$) while the toroidal axisymmetry increases ($60\,^{+28}_{-29}\,\%$ to $95\,^{+2}_{-3}\,\%$), and the radial dipolar axis migrates toward the equator, coinciding with a clear decrease in the latitude of peak radial field intensity in Figure \ref{fig:FractionalMagField1}. These trends are not sensitive to variations in the model parameters within the ranges we have tested (Section \ref{section:Sensitivity analysis}). We also tested a reduced data set for Jan 2016 to balance the number of observations across the two epochs, and this did not impact significantly on the trends we observed between Dec 2015 and Jan 2016. Therefore, we conclude that the observed field evolution is unlikely to be explained by differences in data quantity between these epochs.

\subsubsection{Dec 2016 - Jan 2017}

The Dec 2016 and Jan 2017 data also correspond to a phase of increasing S-index in Figure \ref{fig:FieldGeomSindex}, but the mean field strength decreases and the $B_l$ remains similar for both epochs. This is not surprising as these parameters trace different aspects of magnetic activity that are not necessarily tightly correlated. The mean field strength from ZDI and the $B_l$ reflect fractions of a star's total chromospheric and photospheric magnetic flux, and are associated with the large-scale magnetic structure. Meanwhile, the S-index traces flux variations related to chromospheric plages and network which are smaller, more local features of magnetic activity. Based on Figure \ref{fig:Polarity} both the Dec 2016 and Jan 2017 magnetic maps could coincide with activity cycle minimum. This may explain why we observe minimal changes in the position of the radial dipolar axis and the strength and axisymmetry of the poloidal field between these epochs, in contrast to the trends observed in Oct/Nov 2011 and Dec 2015/Jan 2016. The nature of these subtle changes are sensitive to model parameter variations. Meanwhile, the toroidal axisymmetry shows an increase ($59\,^{+13}_{-27}\,\%$ to $78\,^{+6}_{-9}\,\%$) and this is consistent for all of our sensitivity tests. There is a significant disparity in the number and phase coverage of observations between Jan 2017 and Dec 2016. However, when we tested a reduced data set for Dec 2016 we still observed a decrease in field strength and toroidal axisymmetry from Dec 2016 to Jan 2017, which suggests that the trends are not due to differences in data quantity. 

\subsubsection{Dec 2018 - Jan 2019}

Our Dec 2018 and Jan 2019 observations also appear to correspond with activity cycle minimum in Figure \ref{fig:FieldGeomSindex}, though from Figure \ref{fig:Polarity} they could coincide with a progression between activity cycle maximum and minimum. The mean S-index is the same for both epochs, while the mean field strength increases ($3.1\,^{+0.7}_{-0.8}\,$G to $5.2\,^{+2.8}_{-0.9}\,$G) along with a slight increase in the mean $B_l$ (Figure \ref{fig:Bl}). Although the increase in field strength is not sensitive to changes in the model parameters, we cannot rule out that it may be related to the larger data set available for Jan 2019, which allows for better recovery of the magnetic field. When we tested a reduced data set for Jan 2019 we found the mean field strength to decrease between 2018 and 2019. The strength of the poloidal field is similar in Dec and Jan, but the poloidal field becomes more dipolar. Both the poloidal and toroidal field may also become more axisymmetric. These trends are all sensitive to variations in the adopted model parameters or differences in data quantity between the epochs. The most robust trend is the movement of the radial dipole axis slightly away from the equator between 2018 and 2019; this change is not sensitive to variations in the model parameters within the range we tested, nor is it sensitive to differences in data quantity.

\subsubsection{Comparison to the solar magnetic cycle}

Our results are consistent with HD\,75332 showing a rapid, solar-like magnetic cycle. Between activity cycle minimum and maximum, the Sun's magnetic field increases in strength and becomes more complex  as it approaches a polarity reversal, and the large-scale dipole migrates toward the equator.  Figure \ref{fig:FieldGeomSindex} suggests that polarity reversals may have taken place for HD\,75332 following our Oct 2011/Nov 2011 and Dec 2015/Jan 2016 observations, and if so, the field evolution we have described in each case is distinctly solar-like. Imminent polarity reversals would also explain the emerging positive polarity regions we observed at high latitudes in Nov 2011 and Jan 2016. This may be similar to the possible overlapping butterfly diagram observed for $\tau$\,Boo \citep{Jeffers2018}, where one magnetic cycle begins before the previous cycle has finished. Following a polarity reversal, the structure of the solar magnetic field becomes increasingly poloidal and dipolar as the radial dipolar axis becomes more aligned with the poles. We have described a similar migration of the dipole away from the equator and strengthening of the dipolar component of the poloidal field for HD\,75332 between Dec 2018 and Jan 2019, which may follow a polarity reversal at activity cycle maximum in Figure \ref{fig:Polarity}.

Our observations in Jan-Feb 2007 and Jan 2015 appear to have occurred on the approach to a possible radial polarity reversal according to Figures \ref{fig:Polarity} and \ref{fig:FieldGeomSindex}. The magnetic field in Jan 2015 is strongly toroidal, the radial dipolar axis is close to the equator and we observed a mixed polarity at the visible pole, which are all consistent with an imminent radial polarity reversal.  Conversely, the magnetic structure observed in Jan-Feb 2007 is simple and strongly poloidal ($83\,^{+6}_{-17}\,\%$), and the radial dipole axis is located far from the equator. Given that the polarization signatures and $B_l$ we observed for Nov 2006 show an opposite polarity to that observed in Jan-Feb 2007, it seems likely that the Jan-Feb 2007 observations have captured HD\,75332 following a polarity reversal and in a phase of decreasing activity. Clearly this is at odds with the cycle position shown in Figures \ref{fig:Polarity} and \ref{fig:FieldGeomSindex}. This may be explained by the fact that our data may cover as many as 25 activity cycles, so small errors in our derived cycle period would propagate to much larger errors over many cycles. Accurately placing the observations within cycles may be further complicated by variations in cycle length, which are known to occur in the Sun \citep{Usoskin2003}. 

The magnetic field in Jan 2010 appears to be consistent with its location at activity cycle minimum in Figure \ref{fig:FieldGeomSindex}, being dominantly poloidal with a low field magnitude. We would expect to see a similar, simple structure in Nov-Dec 2017 given the position of the data within the 193.5\,d activity cycle, but the field structure is slightly more complex and significant energy is stored in the toroidal component. Again, this apparent departure from a solar-like regime may be due to changes/errors in cycle length resulting in the inaccurate placement of the data within the activity cycle.

If a 1.06\,yr magnetic cycle is present for HD\,75332, it would suggest that our observations may have `missed' as many as 22 large-scale polarity reversals. This is possible, considering the low frequency of our spectropolarimetric observations. At worst, observational epochs are separated by over 3 years, and at best the observations were carried out annually. If polarity reversals occur every 193.5\,d (0.53\,yr) then there is a high likelihood of two reversals occurring between annual observations, and this would appear as if no reversal had occurred at all. Carefully planned observations would be required to improve the likelihood of directly observing a polarity reversal.

\section[HD75332 vs Tau Boo]{HD\,75332 \MakeLowercase{vs} {$\mathbf{\tau}$} Boo}

$\tau$\,Boo and HD\,75332 have similar physical properties (Table \ref{tab:StellarParam}) and are two of only three F-type stars that have had their magnetic fields and chromospheric activity monitored simultaneously over multiple epochs. Both stars have a similar mass and radius, though HD\,75332 is older and more slowly rotating compared to $\tau$\,Boo. The reported magnetic variability of $\tau$\,Boo is very similar to what we have described here for HD\,75332. $\tau$\,Boo has a weak magnetic field, with mean field strength ranging from 0.9\,G to 3.9\,G across studies \citep{Fares2009,Mengel2016,Jeffers2018} and mean S-index of 0.202 \citep{Wright2004}. It has a rapid $\sim$120\,d chromospheric activity cycle ($\sim$240\,d magnetic cycle), throughout which the large-scale magnetic field evolves in a solar-like fashion. Multiple reversals of the large-scale magnetic field in-phase with chromospheric activity cycles have been observed \citep{Mengel2016,Jeffers2018}. The stable, cyclic magnetic behaviour of both HD\,75332 and $\tau$ Boo contrasts with the magnetic activity observed for younger, more active stars like HD\,171488 (G0V, $M=1.06\pm0.02M_{\odot}$, $R=1.15\pm0.08R_{\odot}$, $P_{rot}\,=\,1.3371\,\pm\,0.0002$\,d), which shows a complex and rapidly varying magnetic field that has not yet been observed to change its polarity \citep{Strassmeier2003,Marsden2006,Jeffers2008,Jeffers2011}. 

A key difference between HD\,75332 and $\tau$ Boo is that $\tau$\,Boo is known to host a HJ whose mass ($\sim$6$\,M_{Jupiter}$) is $\sim$12 times that of the star's thin convective envelope \citep{Fares2009}. \citet{Fares2009} speculated that the HJ may speed up magnetic cycles in $\tau$\,Boo by synchronising with its outer convective envelope and increasing rotational shear. For HD\,75332, no exoplanets have so-far been detected, though \citet{Hinkel2019} predicted a 65\,\% likelihood that HD\,75332 hosts a giant planet ($M\geq0.0945\,M_{Jupiter}$) based on trends in stellar elemental abundances. HD\,75332 has RV$_{RMS}=38.77\textrm{\,m\,s}^{-1}$ \citep*{Fischer2014}, which is much lower compared to the RV semi-amplitude of 469\,m\,s$^{-1}$ for $\tau$\,Boo \citep{Butler1997}. Since HD\,75332 appears to have a similar magnetic cycle to $\tau$\,Boo but no HJ, it is unlikely that the rapid cycle observed for $\tau$\,Boo can be attributed to star-planet interaction. Rapid cycles may be intrinsic to late-F stars and possibly related to their shallow outer convective envelopes. 

Like $\tau$\,Boo, HD\,75332 is capable of generating a strong toroidal field, which \citet{See2016} suggests might be a common feature among stars on the `active sequence' in the activity cycle period ($P_{\textrm{cyc}}$) / rotation period plane. Active sequence stars also typically have two independent chromospheric activity cycles; one corresponding to each of the active and inactive sequences \citep{See2016,Brandenburg2017}. These may correspond to two different dynamo mechanisms; for stars on the inactive sequence the dominant dynamo is that operating at the tachocline, while for active sequence stars it is a dynamo operating in the near-surface shear layers \citep{Bohm2007,Brandenburg2017}. The existence of such activity sequences has recently been questioned \citep{BoroSaikia2018} but it is interesting that our detected 193.5\,d ($\log(P_{\textrm{rot}}/P_{\textrm{cyc}})=-1.74$) and 3.9\,yr ($\log(P_{\textrm{rot}}/P_{\textrm{cyc}})=-2.60$)  S-index cycles lie roughly on the empirical inactive and active branches from \citet{Brandenburg2017}, based on $\log{R'_{HK}}$=-4.51 from \citet{marsden2014}. The fact that we have detected three independent chromospheric activity cycles for HD\,75332 is also interesting, as this cannot be explained by two dynamo modes. Two chromospheric activity cycles of 11.6\,yr \citep{Baliunas1997} and $\sim$120\,d \citep{Mengel2016} have been identified for $\tau$\,Boo, though \citet{Mengel2016} also detected possible cycles with periods between $\sim$300 and several thousand days.  \citet{See2016} speculates that HD\,78366 (F9V), the only other F star that has had its large-scale surface magnetic field mapped over multiple epochs,  may also have three cycles. \citet{Baliunas1995} detected S-index cycles of 12.2\,yr and 5.9\,yr, while \citet{Morgenthaler2011} observed two large-scale magnetic polarity reversals that indicated a $\leq3$\,yr magnetic cycle. Follow-up observations of HD\,75332, $\tau$\,Boo and HD\,78366 would be interesting to determine whether or not they have three chromospheric activity cycles, with one corresponding to the magnetic cycle. 
\section{Conclusions}

In this study we combined long-term chromospheric activity data with multi-epoch spectropolarimetric observations to characterize the chromospheric and magnetic variability of the F7V star HD\,75332. We found three significant cycles present in the chromospheric S-index; a $\sim$193.5\,d cycle consistent with the rapid activity cycle detected by \citet{Mittag2019}, as well as $\sim$3.9\,yr and $\sim$31.5\,yr cycles.  From $\sim$13 years of spectropolarimetric data we detected a 377.6\,d periodicity in the longitudinal magnetic field strength. We also reconstructed the large-scale surface magnetic field of HD\,75332 for 12 epochs between Jan-Feb 2007 and Jan 2019. We observed one reversal of the large-scale radial field between Jan 2010 and Oct 2011, but we expect that large-scale polarity reversals may be much more frequent for HD\,75332 based on its rapid, cyclic magnetic variability. The reconstructed field polarity is consistent at all epochs with reversals occurring on a 193.5\,d time-scale, in phase with our detected 193.5\,d chromospheric activity cycle. Between expected polarity reversals the magnetic field appears to vary in a solar-like fashion. When approaching a polarity reversal at activity cycle maximum, the magnetic field grows stronger and more toroidal, while the large-scale radial dipolar axis migrates toward the equator.  Following a polarity reversal, the large-scale radial dipolar axis becomes more aligned with the poles. Based on temporal shifts of magnetic features, we measured a potentially high level of surface differential rotation for HD\,75332, with $d\Omega=0.25\,^{+0.23}_{-0.20}\ \textrm{\,rad\,d}^{-1}$. Varying $d\Omega$ and other model parameters within a large domain did not impact significantly on the polarity of the reconstructed large-scale magnetic field, nor the rapid nature of the magnetic field evolution we observed across epochs. 

Our results indicate that HD\,75332 has a $\sim$1.06\,yr solar-like magnetic cycle, though additional high-cadence observations throughout an entire activity cycle will be required to confirm its presence. This would make it the third star apart from our Sun that is known to exhibit solar-like magnetic behaviour, and the second late-F star to have a rapid, solar-like magnetic cycle. Rapid magnetic cycles may be common among late-F stars and are possibly related to their thin outer convective envelopes.  This suggests that the rapid magnetic cycle observed for $\tau$ Boo is intrinsic and that the dynamo is not sped up by star-planet interaction. If rapid magnetic cycles are common among late-F stars, then this would make them ideal targets to study solar-like magnetic variability within a short time-frame and provide important constraints for solar and stellar dynamo models.

\section*{Acknowledgements}

The authors would like to thank an anonymous referee for their time and very constructive comments which have helped to improve this manuscript.

This work is based on observations obtained with NARVAL at the Télescope Bernard Lyot (TBL), which are operated by INSU/CNRS. We thank the staff at the TBL for their time and data. We also thank the team at the La Luz Observatory, Guanajuato, Mexico for the TIGRE data used in this work. The HK Project data (HK\_Project\_v2001\_NSO) derive from the Mount Wilson Observatory HK Project, which was supported by both public and private funds through the Carnegie Observatories, the Mount Wilson Institute, and the Harvard-Smithsonian Center for Astrophysics, starting in 1966 and continuing for over 36 years.  These data are the result of the dedicated work of O. Wilson, A. Vaughan, G. Preston, D. Duncan, S. Baliunas, and many others. We acknowledge the use of data from the TESS mission, funding for which is provided by NASA’s Science Mission directorate. We also acknowledge use of the SIMBAD and VizieR data bases operated at CDS, Strasbourg, France. 

ELB is supported by an Australian Government Research Training Program (RTP) Scholarship.

SJ acknowledges the support of the  German Science Foundation (DFG) Research Unit FOR2544 "Blue Planets around Red Stars" (project JE 701/3-1) and DFG priority program SPP 1992 "Exploring the Diversity of Extrasolar Planets" (JE 701/5-1).

AAV acknowledges funding from the European Research Council (ERC) under the European Union's Horizon 2020 research and innovation programme (grant agreement No 817540, ASTROFLOW).

VS acknowledges funding from the European Research Council (ERC) under the European Unions Horizon 2020 research and innovation programme (grant agreement No. 682393 AWESoMeStars).

\section*{Data availability}

All NARVAL data presented here are publicly available through the PolarBase data base (http://polarbase.irap.omp.eu/). MWO data are available via the National Solar Observatory website (https://nso.edu/data/historical-data/mount-wilson-observatory-hk-project/) and TESS data can be accessed through MAST (https://mast.stsci.edu/portal/Mashup/Clients/Mast/Portal.html). The TIGRE data presented in this work may be requested by contacting MM (mmittag@hs.uni-hamburg.de).


\bibliographystyle{mnras}

\appendix

\section{Journal of NARVAL observations, sample Stokes {\it{I}} line profile and full series of Stokes {\it{V}} line profiles}
\begin{table*}
\caption{NARVAL observation details. Each Stokes {\it{V}} observation listed here was derived from a series of four Stokes {\it{I}} sub-exposures. For each Stokes {\it{V}} observation, the columns show (left to right) the local date (PolarBase spectra filenames correspond to the local date), HJD (+2454000), total exposure time (s) and Universal Time (UT) at which the observation was taken, along with the rotational cycle with respect to the ephemeris $HJD=2454054.67411+3.56\phi$, rotational cycle with respect to the epoch shown in Table \ref{tab:NARVALdata} ($\phi$), mean SNR, detection of the polarization signal (D:definite, M:marginal, N:non-detection), mean S-index and mean longitudinal magnetic field strength ($B_l$). }
\label{tab:NARVALobsdetails}
\begin{tabular}{cccccccccc}
\toprule
\multirow{2}{*}{\textbf{Date}}      & \textbf{HJD} & \textbf{Exposure} & \multirow{2}{*}{\textbf{UT}} & \multirow{2}{*}{\textbf{Cycle}}  & \multirow{2}{*}{\textbf{$\phi$}} & \multirow{2}{*}{\textbf{Mean SNR}} & \multirow{2}{*}{\textbf{D/M/N}} & \multirow{2}{*}{\textbf{S-index}} & \multirow{2}{*}{\textbf{$B_l$ (G)}}\\ 
  &  \textbf{(+2454000)} & \textbf{time (s)} &   & & & & \\\midrule
14-Nov-06 & 54.67   & 4 x 300 & 15-Nov-06 04:08:46 &0.00 &-2.00  & 22671 & D & 0.275 &   -0.4	$\pm$	1.8\\
29-Nov-06 & 69.73   & 4 x 300 & 30-Nov-06 05:32:51 &4.23 & 2.23  & 20397 & D & 0.287 &   -2.9	$\pm$	2.0\\ \midrule
26-Jan-07 & 127.53  & 4 x 300 & 27-Jan-07 00:34:24 &20.46 &-1.54  & 15455 & D & 0.286 &   6.0	$\pm$	2.6\\
27-Jan-07 & 128.54  & 4 x 300 & 28-Jan-07 00:49:52 &20.75 &-1.25  & 20788 & M & 0.281 &   -0.3	$\pm$	1.9\\
29-Jan-07 & 130.53  & 4 x 300 & 30-Jan-07 00:42:23 &21.31 &-0.69  & 23574 & D & 0.273 &   3.3	$\pm$	1.7\\
1-Feb-07  & 133.56  & 4 x 300 & 2-Feb-07 01:19:00 & 22.16&  0.16 & 17307 & D & 0.270 &   3.9	$\pm$	2.3\\
2-Feb-07  & 134.53  & 4 x 300 & 3-Feb-07 00:36:30 & 22.43&  0.43 & 19097 & N & 0.275 &   1.4	$\pm$	2.1\\
3-Feb-07  & 135.55  & 4 x 300 & 4-Feb-07 01:07:35 & 22.72&  0.72 & 21726 & D & 0.281 &   8.2	$\pm$	1.8\\
4-Feb-07  & 136.52  & 4 x 300 & 5-Feb-07 00:26:24 & 22.99&  0.99 & 17240 & N & 0.278 &   6.0	$\pm$	2.4\\
8-Feb-07  & 140.53  & 4 x 300 & 9-Feb-07 00:43:38 & 24.12&  2.12 & 3135  & N & 0.273 &   1.7	$\pm$	12.9\\ \midrule
30-Jan-09 & 862.61  & 4 x 400 & 31-Jan-09 02:28:43 &226.95 &-0.05  & 22099 & D & 0.264 &   7.2	$\pm$	1.8\\ \midrule
14-Dec-09 & 1180.73 & 4 x 300 & 15-Dec-09 05:28:55 &316.31 & 0.31  & 6842  & N & 0.268 &   4.0	$\pm$	5.8\\
15-Dec-09 & 1181.71 & 4 x 300 & 16-Dec-09 04:54:28 &316.58 & 0.58  & 22323 & N & 0.270 &   -0.8	$\pm$	1.8\\ \midrule
5-Jan-10  & 1202.67 & 4 x 300 & 6-Jan-10 03:55:28 & 322.47&  0.47 & 17833 & M & 0.269 &   0.5	$\pm$	2.2\\ \midrule
18-Jan-10 & 1215.60 & 4 x 300 & 19-Jan-10 02:13:26 &326.10 &-0.90  & 14503 & N & 0.281 &   -2.8	$\pm$	2.8\\
22-Jan-10 & 1219.47 & 4 x 300 & 22-Jan-10 23:12:02 &327.19 & 0.19  & 15831 & M & 0.262 &   1.3	$\pm$	2.5\\
25-Jan-10 & 1222.56 & 4 x 300 & 26-Jan-10 01:21:27 &328.06 & 1.06  & 20506 & D & 0.263 &   3.6	$\pm$	2.0\\
27-Jan-10 & 1224.58 & 4 x 300 & 28-Jan-10 01:56:40 &328.63 & 1.63  & 20420 & M & 0.265 &   2.1	$\pm$	2.0\\ \midrule
13-Feb-10 & 1241.56 & 4 x 300 & 14-Feb-10 01:26:23 &333.40 &-0.60  & 12140 & N & 0.264 &   3.3	$\pm$	3.3\\
14-Feb-10 & 1242.54 & 4 x 300 & 15-Feb-10 00:55:08 &333.67 &-0.33  & 15013 & N & 0.262 &   -1.8	$\pm$	2.7\\
15-Feb-10 & 1243.53 & 4 x 300 & 16-Feb-10 00:33:58 &333.95 &-0.05  & 13739 & N & 0.260 &   1.0	$\pm$	2.9\\ \midrule
4-Oct-11  & 1839.71 & 4 x 300 & 5-Oct-11 05:00:14 & 501.41& -1.59 & 18168 & N & 0.270 &   -3.2	$\pm$	2.2\\
8-Oct-11  & 1843.64 & 4 x 300 & 9-Oct-11 03:17:48 & 502.52& -0.48 & 15794 & N & 0.269 &   -7.0	$\pm$	2.5\\
11-Oct-11 & 1846.67 & 4 x 300 & 12-Oct-11 04:02:33 &503.37 & 0.37  & 9643  & N & 0.266 &   -8.3	$\pm$	4.2\\
13-Oct-11 & 1848.68 & 4 x 300 & 14-Oct-11 04:16:24 &503.93 & 0.93  & 22420 & D & 0.264 &   -3.4	$\pm$	1.8\\ \midrule
8-Nov-11  & 1874.70 & 4 x 300 & 9-Nov-11 04:47:34 & 511.24& -2.76 & 13704 & N & 0.264 &   -2.0	$\pm$	2.9\\
10-Nov-11 & 1876.66 & 4 x 300 & 11-Nov-11 03:49:17 &511.79 &-2.21  & 19793 & N & 0.276 &   -0.2	$\pm$	2.0\\
11-Nov-11 & 1877.74 & 4 x 300 & 12-Nov-11 05:40:23 &512.10 &-1.90  & 16519 & D & 0.268 &   -4.7	$\pm$	2.4\\
23-Nov-11 & 1889.70 & 4 x 300 & 24-Nov-11 04:50:59 &515.46 & 1.46  & 15946 & D & 0.278 &   -7.3	$\pm$	2.5\\
25-Nov-11 & 1891.73 & 4 x 300 & 26-Nov-11 05:24:49 &516.03 & 2.03  & 19504 & D & 0.283 &   -3.8	$\pm$	2.0\\
26-Nov-11 & 1892.70 & 4 x 300 & 27-Nov-11 04:50:08 &516.30 & 2.30  & 14687 & D & 0.281 &   -8.0	$\pm$	2.7\\ \midrule
5-Jan-15  & 3028.63 & 4 x 300 & 6-Jan-15 03:07:58 & 835.38& -1.62 & 19646 & D & 0.272 &   -5.4	$\pm$	2.0\\
6-Jan-15  & 3029.62 & 4 x 300 & 7-Jan-15 02:52:17 & 835.66& -1.34 & 20894 & D & 0.272 &   -1.3	$\pm$	1.9\\
7-Jan-15  & 3030.56 & 4 x 300 & 8-Jan-15 01:25:29 & 835.92& -1.08 & 19501 & D & 0.278 &   -4.0	$\pm$	2.0\\
8-Jan-15  & 3031.56 & 4 x 300 & 9-Jan-15 01:21:30 & 836.20& -0.80 & 16081 & N & 0.277 &   -5.5	$\pm$	2.5\\
9-Jan-15  & 3032.58 & 4 x 300 & 10-Jan-15 01:50:08 &836.49 &-0.51  & 20743 & D & 0.273 &   -1.7	$\pm$	1.9\\
10-Jan-15 & 3033.59 & 4 x 300 & 11-Jan-15 02:02:35 &836.77 &-0.23  & 21057 & D & 0.277 &   -1.1	$\pm$	1.9\\
11-Jan-15 & 3034.60 & 4 x 300 & 12-Jan-15 02:24:06 &837.06 & 0.06  & 18812 & D & 0.279 &   -0.4	$\pm$	2.1\\
17-Jan-15 & 3040.56 & 4 x 300 & 18-Jan-15 01:20:06 &838.73 & 1.73  & 12131 & D & 0.272 &   -2.9	$\pm$	3.3\\ \midrule
1-Dec-15  & 3358.63 & 4 x 300 & 2-Dec-15 02:57:28 & 928.08& -1.92 & 14979 & D & 0.254 &   -2.1	$\pm$	2.7\\
3-Dec-15  & 3360.70 & 4 x 300 & 4-Dec-15 04:41:24 & 928.66& -1.34 & 13067 & N & 0.261 &   -5.0	$\pm$	3.1\\
12-Dec-15 & 3369.58 & 4 x 300 & 13-Dec-15 01:54:54 &931.15 & 1.15  & 13836 & D & 0.257 &   -4.0	$\pm$	2.9\\
17-Dec-15 & 3374.62 & 4 x 300 & 18-Dec-15 02:52:57 &932.57 & 2.57  & 4824  & N & 0.262 &   -17.2	$\pm$	8.4\\ \midrule
20-Jan-16 & 3408.72 & 4 x 300 & 21-Jan-16 05:06:33 &942.15 &-0.85  & 16893 & N & 0.262 &   -3.7	$\pm$	2.4\\
21-Jan-16 & 3409.65 & 4 x 300 & 22-Jan-16 03:25:14 &942.41 &-0.59  & 14324 & D & 0.260 &   -9.3	$\pm$	2.8\\
24-Jan-16 & 3412.70 & 4 x 300 & 25-Jan-16 04:38:09 &943.26 & 0.26  & 14980 & D & 0.262 &   -5.7	$\pm$	2.7\\
25-Jan-16 & 3413.67 & 4 x 300 & 26-Jan-16 03:53:16 &943.54 & 0.54  & 9185  & M & 0.266 &   -3.3	$\pm$	4.4\\
26-Jan-16 & 3414.70 & 4 x 300 & 27-Jan-16 04:39:34 &943.83 & 0.83  & 1063   & N & 0.278&   3.3	$\pm$	37.6 \\
27-Jan-16 & 3415.46 & 4 x 300 & 27-Jan-16 23:01:42 &944.04 & 1.04  & 16426 & D & 0.279 &   -1.3	$\pm$	2.5\\
29-Jan-16 & 3417.57 & 4 x 300 & 30-Jan-16 01:33:47 &944.63 & 1.63  & 14182 & D & 0.268 &   -1.8	$\pm$	2.8\\ \midrule
28-Oct-16 & 3690.60 & 4 x 300 & 29-Oct-16 02:18:42 &1021.33 & 0.33  & 10290 & M & 0.263 &   -1.0	$\pm$	3.9\\ \bottomrule
\end{tabular}
\end{table*}

\begin{table*}
\contcaption{}
\label{tab:continued}
\begin{tabular}{cccccccccc}
\toprule
\multirow{2}{*}{\textbf{Date}}      & \textbf{HJD} & \textbf{Exposure} & \multirow{2}{*}{\textbf{UT}} & \multirow{2}{*}{\textbf{Cycle}}  & \multirow{2}{*}{\textbf{$\phi$}} & \multirow{2}{*}{\textbf{Mean SNR}} & \multirow{2}{*}{\textbf{D/M/N}} & \multirow{2}{*}{\textbf{S-index}} & \multirow{2}{*}{\textbf{$B_l$ (G)}}\\ 

  &  \textbf{(+2454000)} & \textbf{time (s)} &   & & & &  \\\midrule
2-Dec-16  & 3725.68 & 4 x 300 & 3-Dec-16 04:18:14 & 1031.18&  -2.82 & 18338 & D & 0.265  &   -2.1	$\pm$	2.2\\
6-Dec-16  & 3729.69 & 4 x 300 & 7-Dec-16 04:35:21 & 1032.31&  -1.69 & 17492 & D & 0.268  &   -3.6	$\pm$	2.3\\
7-Dec-16  & 3730.69 & 4 x 300 & 8-Dec-16 04:25:29 & 1032.59&  -1.41 & 19338 & N & 0.260  &   -4.8	$\pm$	2.1\\
8-Dec-16  & 3731.69 & 4 x 300 & 9-Dec-16 04:35:21 & 1032.87&  -1.13 & 19342 & N & 0.262  &   -5.0	$\pm$	2.1\\
9-Dec-16  & 3732.69 & 4 x 300 & 10-Dec-16 04:26:59 &1033.15 & -0.85  & 16152 & D & 0.261  &   -5.9	$\pm$	2.5\\
10-Dec-16 & 3733.66 & 4 x 300 & 11-Dec-16 03:47:34 &1033.42 & -0.58  & 17644 & N & 0.263  &   -4.2	$\pm$	2.3\\
12-Dec-16 & 3735.67 & 4 x 300 & 13-Dec-16 03:58:52 &1033.99 & -0.01  & 18140 & D & 0.261  &   -8.8	$\pm$	2.2\\
13-Dec-16 & 3736.70 & 4 x 300 & 14-Dec-16 04:40:38 &1034.28 &  0.28  & 16930 & D & 0.260  &   -4.0	$\pm$	2.4\\
14-Dec-16 & 3737.69 & 4 x 300 & 15-Dec-16 04:36:27 &1034.56 &  0.56  & 18763 & N & 0.257  &   -5.9	$\pm$	2.2\\
17-Dec-16 & 3740.75 & 4 x 300 & 18-Dec-16 05:52:53 &1035.41 &  1.41  & 12688 & M & 0.271  &   -3.7	$\pm$	3.2\\
18-Dec-16 & 3741.70 & 4 x 300 & 19-Dec-16 04:45:55 &1035.68 &  1.68  & 15747 & N & 0.267  &   -0.3	$\pm$	2.6\\ \midrule
2-Jan-17  & 3756.59 & 4 x 300 & 3-Jan-17 02:10:14 & 1039.87&  -1.13 & 7910  & N & 0.267  &   -9.0	$\pm$	5.2\\
3-Jan-17  & 3757.68 & 4 x 300 & 4-Jan-17 04:20:38 & 1040.17&  -0.83 & 18384 & N & 0.273  &   -1.5	$\pm$	2.2\\
6-Jan-17  & 3760.63 & 4 x 300 & 7-Jan-17 02:55:45 & 1041.00&  -0.00 & 17704 & D & 0.272  &   -2.8	$\pm$	2.3\\
7-Jan-17  & 3761.65 & 4 x 300 & 8-Jan-17 03:29:05 & 1041.29&   0.29 & 17397 & M & 0.268  &   -5.4	$\pm$	2.3\\ \midrule
1-Nov-17  & 4059.57 & 4 x 300 & 2-Nov-17 01:44:21 & 1124.97&  -0.03 & 12796 & N & 0.285  &   1.0	$\pm$	3.2\\ \midrule
15-Nov-17 & 4073.67 & 4 x 300 & 16-Nov-17 03:56:38 &1128.93 & -2.07  & 15274 & M & 0.275  &   -1.1	$\pm$	2.7\\
16-Nov-17 & 4074.68 & 4 x 300 & 17-Nov-17 04:14:04 &1129.21 & -1.79  & 13592 & N & 0.283  &   1.4	$\pm$	3.0\\
17-Nov-17 & 4075.65 & 4 x 300 & 18-Nov-17 03:39:28 &1129.49 & -1.51  & 17023 & D & 0.279  &   -2.1	$\pm$	2.4\\
18-Nov-17 & 4076.72 & 4 x 300 & 19-Nov-17 05:08:13 &1129.79 & -1.21  & 16627 & N & 0.273  &   2.1	$\pm$	2.4\\
19-Nov-17 & 4077.66 & 4 x 300 & 20-Nov-17 03:54:13 &1130.05 & -0.95  & 16937 & D & 0.277  &   -2.3	$\pm$	2.4\\
20-Nov-17 & 4078.66 & 4 x 300 & 21-Nov-17 03:44:36 &1130.33 & -0.67  & 13805 & M & 0.283  &   1.0	$\pm$	2.9\\
21-Nov-17 & 4079.71 & 4 x 300 & 22-Nov-17 05:06:13 &1130.63 & -0.37  & 11459 & N & 0.283  &   3.6	$\pm$	3.5\\
26-Nov-17 & 4084.72 & 4 x 300 & 27-Nov-17 05:18:16 &1132.04 &  1.04  & 12084 & N & 0.277  &   0.8	$\pm$	3.4\\
27-Nov-17 & 4085.74 & 4 x 300 & 28-Nov-17 05:37:05 &1132.32 &  1.32  & 16635 & N & 0.281  &   -0.2	$\pm$	2.4\\
4-Dec-17  & 4092.67 & 4 x 300 & 5-Dec-17 03:56:19 & 1134.27&   3.27 & 13118 & D & 0.273  &   -1.4	$\pm$	3.1\\
5-Dec-17  & 4093.62 & 4 x 300 & 6-Dec-17 02:47:51 & 1134.54&   3.54 & 12300 & D & 0.279  &   -1.3	$\pm$	3.3\\ \midrule
23-Jan-18 & 4142.73 & 4 x 300 & 24-Jan-18 05:27:38 &1148.33 &  0.33  & 7137  & N & 0.269  &   -4.7	$\pm$	5.7\\ \midrule
14-Nov-18 & 4437.68 & 4 x 300 & 15-Nov-18 04:16:57 &1231.18 &  0.18  & 18581 & N & 0.277  &   -4.4	$\pm$	2.2\\
16-Nov-18 & 4439.63 & 4 x 300 & 17-Nov-18 03:10:05 &1231.73 &  0.73  & 18582 & D & 0.287  &   -5.0	$\pm$	2.2\\ \midrule
4-Dec-18  & 4457.72 & 4 x 300 & 5-Dec-18 05:12:49 & 1236.81&  -1.19 & 16888 & M & 0.280 &   -0.2	$\pm$	2.4 \\
6-Dec-18  & 4459.65 & 4 x 300 & 7-Dec-18 03:32:23 & 1237.35&  -0.65  & 13429 & N & 0.272  &   -6.2	$\pm$	3.1\\
10-Dec-18 & 4463.72 & 4 x 300 & 11-Dec-18 05:12:40 &1238.50 &  0.50 & 20021 & D & 0.270  &   -4.4	$\pm$	2.1\\
17-Dec-18 & 4470.65 & 4 x 300 & 18-Dec-18 03:30:34 &1240.44 &  2.44 & 9474  & M & 0.280  &   -4.5	$\pm$	4.3\\ \midrule
3-Jan-19  & 4487.64 & 4 x 300 & 4-Jan-19 03:19:11 & 1245.22&  -2.78 & 17119 & D & 0.281  &   -2.1	$\pm$	2.4\\
4-Jan-19  & 4488.65 & 4 x 300 & 5-Jan-19 03:35:06 & 1245.50&  -2.50 & 17123 & N & 0.278  &   -3.2	$\pm$	2.4\\
5-Jan-19  & 4489.51 & 4 x 300 & 5-Jan-19 24:07:11 & 1245.74&  -2.26 & 11719 & M & 0.270  &   -2.1	$\pm$	3.5\\
6-Jan-19  & 4490.52 & 4 x 300 & 7-Jan-19 00:26:53 & 1246.02&  -1.98 & 16825 & D & 0.274  &   -6.9	$\pm$	2.4\\
7-Jan-19  & 4491.68 & 4 x 300 & 8-Jan-19 04:17:19 & 1246.35&  -1.65 & 17244 & D & 0.280  &   -2.7	$\pm$	2.4\\
12-Jan-19 & 4496.65 & 4 x 300 & 13-Jan-19 03:37:50 &1247.75 & -0.25  & 13886 & M & 0.269  &   -1.3	$\pm$	3.0\\
15-Jan-19 & 4499.48 & 4 x 300 & 15-Jan-19 23:31:32 &1248.54 &  0.54  & 12598 & N & 0.275  &   1.0	$\pm$	3.2\\
16-Jan-19 & 4500.66 & 4 x 300 & 17-Jan-19 03:45:42 &1248.87 &  0.87  & 15723 & D & 0.269  &   0.5	$\pm$	2.6\\
21-Jan-19 & 4505.66 & 4 x 300 & 22-Jan-19 03:38:49 &1250.28 &  2.28  & 8441  & N & 0.272  &   3.7	$\pm$	4.8\\
26-Jan-19 & 4510.60 & 4 x 300 & 27-Jan-19 02:25:10 &1251.67 &  3.67  & 17485 & D & 0.279  &   -7.5	$\pm$	2.3\\
 \bottomrule
\end{tabular}
\end{table*}

\begin{figure}
    \centering
    \includegraphics[width=0.8\columnwidth]{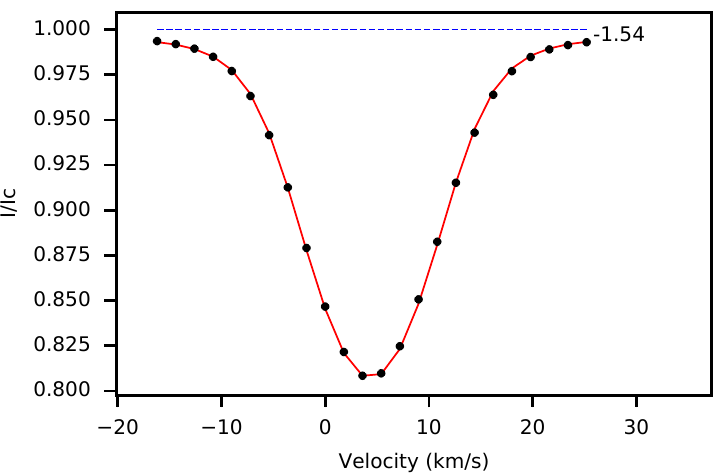}
    \caption{Sample Stokes {\it{I}} model line profile (red) and observed Stokes {\it{I}} LSD profile (black) for 26 Jan 2007. The continuum level is shown by the blue dashed line. $\pm\,1\sigma$ error bars are included for each data point but they are smaller than the symbol size. The rotational cycle of the observation with respect to the epoch in Table \ref{tab:NARVALdata} is shown on the right of the profiles and assumes $P_{rot}=3.56\textrm{\,d}$.}
    \label{fig:DI_fit}
\end{figure}

\begin{figure*}
\includegraphics[width=\linewidth]{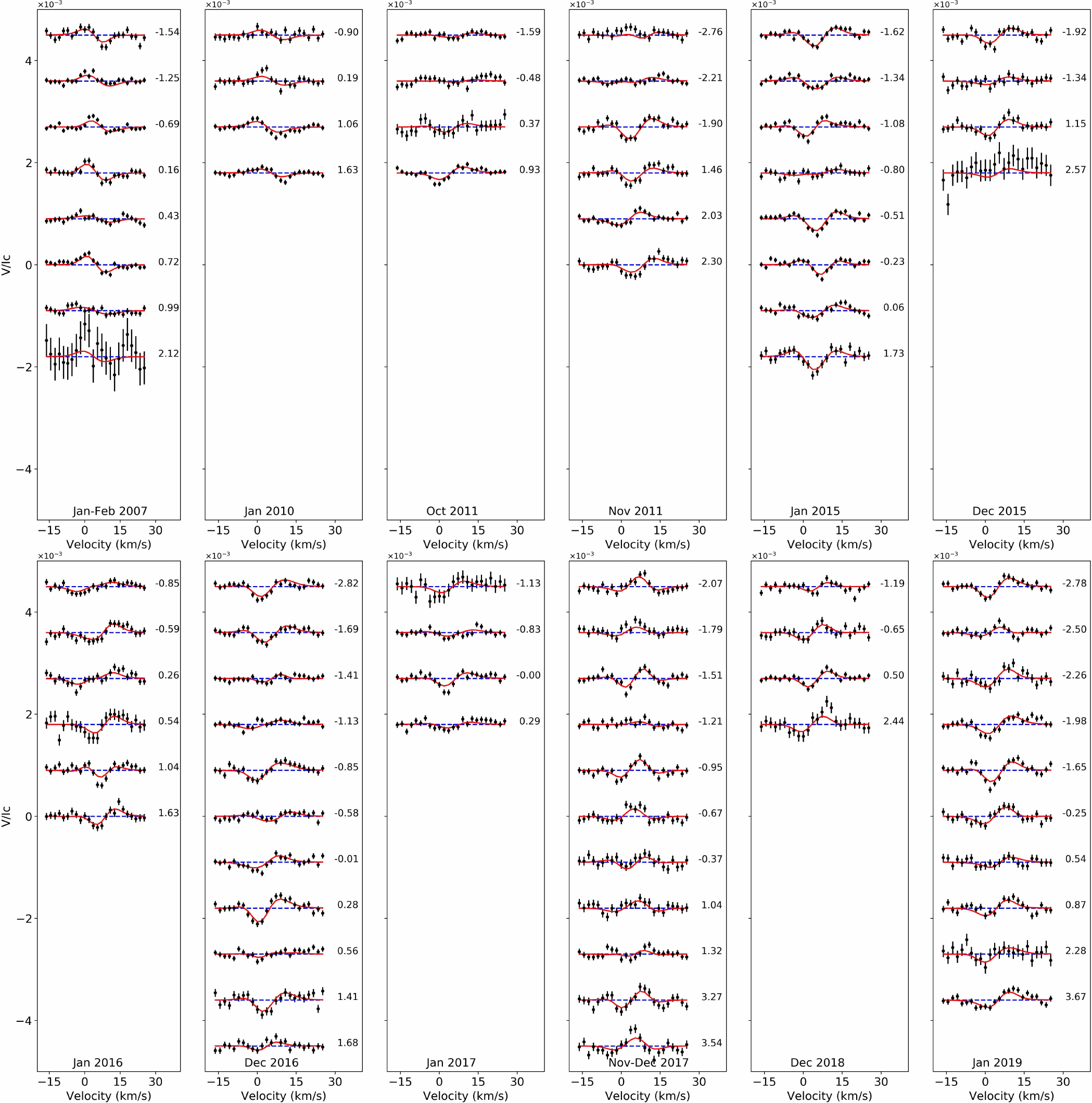}
\caption{Model Stokes {\it{V}} profiles (red) and observed Stokes {\it{V}} LSD profiles (black) for Jan-Feb 2007 through to Jan 2019. Error bars shown are $\pm\,1\sigma$. The rotational cycles of each observation  with respect to the epochs in Table \ref{tab:NARVALdata} are shown on the right of the profiles and assume $P_{rot}=3.56\textrm{\,d}$. Note: For clarity we did not include a profile for 26 Jan 2016 due to the very low SNR, which caused a scattering of data across the chart.}
\label{fig:ZDIfits}
\end{figure*}

\begin{figure}
\centering
\includegraphics[width=0.7\columnwidth]{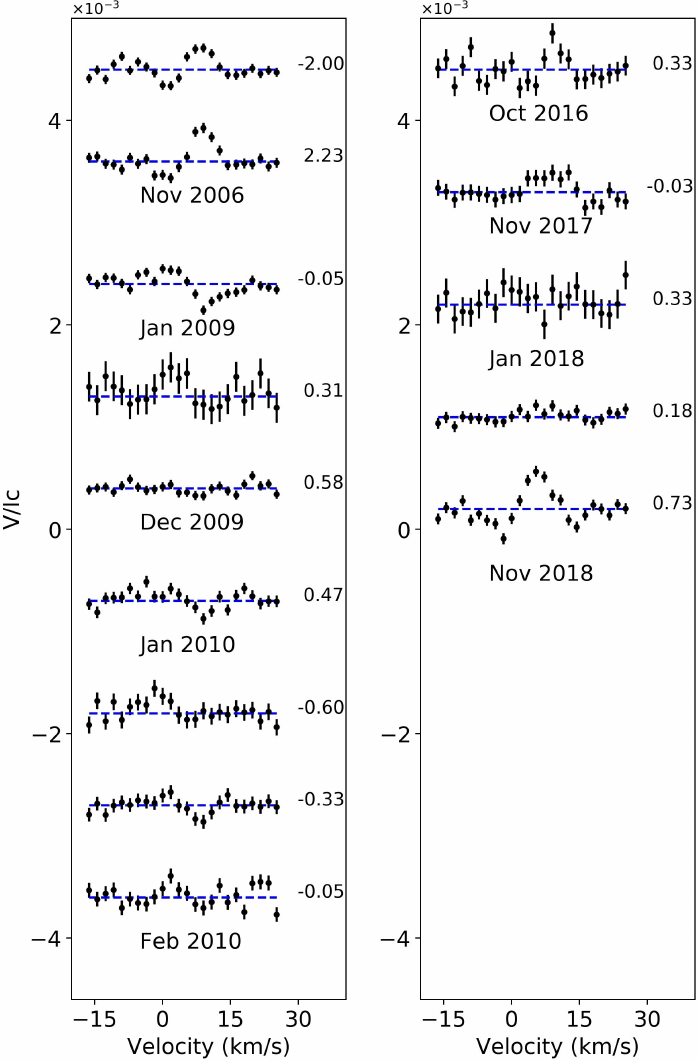}
\caption{Observed Stokes {\it{V}} LSD profiles for small data sets. Error bars shown are $\pm\,1\sigma$. The rotational cycles of each observation  with respect to the epochs in Table \ref{tab:NARVALdata} are shown on the right of the profiles and assume $P_{rot}=3.56\textrm{\,d}$.}
\label{fig:ZDIfits2}
\end{figure}


\bsp	
\label{lastpage}
\end{document}